\def\clearAllMarkings{1}

\ifdefined\icml
\documentclass{article}
\usepackage{fontspec}
\usepackage{siunitx}
\fi

\documentclass[journal]{IEEEtran}
\IEEEoverridecommandlockouts

\usepackage{algorithm}
\usepackage{algorithmic}

\usepackage{caption}
\usepackage{ulem}

\usepackage[pdftex,dvipsnames]{xcolor}  
\usepackage{xargs}  
\usepackage[colorinlistoftodos,prependcaption,textsize=tiny]{todonotes}
\usepackage{microtype}
\usepackage{array,multirow,graphicx}
\newcolumntype{H}{>{\setbox0=\hbox\bgroup}c<{\egroup}@{}}
\graphicspath{{.},{./figs/}}
\usepackage{subfigure}
\usepackage{booktabs} 
\usepackage{float}
\usepackage{amssymb}

\usepackage{makecell}
\usepackage{adjustbox}
\usepackage{xspace}
\usepackage[acronym]{glossaries}
\usepackage{python}
\usepackage{csvsimple}
\usepackage{textgreek}
\usepackage{pgfplots}
\pgfplotsset{compat=1.13}
\usepackage{pgfplotstable}

\usepackage{footmisc}
\usepackage[sort&compress, numbers]{natbib}
\usepackage{pagecolor}
\usepackage{tablefootnote}
\usepackage{atbegshi} 

\usepackage{tikz-inet}
\usetikzlibrary{matrix}
\usepackage{tikz}
\usepackage{mathtools}
\usepackage{multirow}
\usepackage{hyperref} 

\newcommand{\figures}[1]{./#1}

\ifdefined\icml
\usepackage[accepted]{icml2018}
\fi
\usepackage{microtype,xparse,tcolorbox}
\newenvironment{reviewer-comment }{}{}
\tcbuselibrary{skins}
\tcolorboxenvironment{reviewer-comment }{empty,
  left = 1em, top = 1ex, bottom = 1ex,
  borderline west = {2pt} {0pt} {black!20},
}
\ExplSyntaxOn
\NewDocumentEnvironment {response} { +m O{black!20} } {
  \IfValueT {#1} {
    \begin{reviewer-comment~}
      \setlength\parindent{2em}
      \noindent
      \ttfamily #1
    \end{reviewer-comment~}
  }
  \par\noindent\ignorespaces
} { \bigskip\par }

\ExplSyntaxOff

\DeclarePairedDelimiter\floor{\lfloor}{\rfloor}
\newcommand{\calc}[2]{\FPeval{\calcresult}{round(#2,#1)}\calcresult}
\newcommand{\x}{$\times$}
\newcommand{\rmv}[1]{}
\newcommand{\tableSoaSpacing}{\hskip 2mm}

\def\kilo{1000}
\def\mega{1000}
\def\mega{1000000}
\def\giga{1000000000}

\def\tokmg#1{
    \ifnum#1>\giga
      \FPeval{\calcresult}{round(#1/\giga,0)}\calcresult G
    \else
        \ifnum#1>\mega
          \FPeval{\calcresult}{round(#1/\mega,0)}\calcresult M
        \else
            \ifnum#1>\kilo
              \FPeval{\calcresult}{round(#1/\kilo,0)}\calcresult k
            \else
              #1
            \fi
        \fi
    \fi
}

\newcommandx{\info}[2][1=]{\todo[linecolor=OliveGreen,backgroundcolor=OliveGreen!25,bordercolor=OliveGreen,#1]{#2}}
\newcommandx{\unsure}[2][1=]{\todo[linecolor=red,backgroundcolor=red!25,bordercolor=red,#1]{#2}}
\newcommandx{\change}[2][1=]{\todo[linecolor=blue,backgroundcolor=blue!25,bordercolor=blue,#1]{#2}}
\newcommandx{\improvement}[2][1=]{\todo[linecolor=Plum,backgroundcolor=Plum!25,bordercolor=Plum,#1]{#2}}
\newcommand{\hide}[1]{}

\newcommand{\rev}[1]{#1}

 \newcommand{\rv}[1]{#1}
  \newcommand{\jetcas}[1]{#1}
  \newcommand{\jetcasIF}[2]{}

 \newcommand{\revTWO}[1]{{#1}}
 
     \newcommand{\revONEnotbold}[1]{{#1}}
  \newcommand{\revONE}[1]{{#1}}

 \ifdefined\clearAllMarkings
 \newcommand{\TBC}[1]{#1}
    
  \newcommand{\markblue}[1]{#1}
  \newcommand{\todoh}[2]{}
  
 \newcommand{\jetcasVB}[1]{{{#1}}}
 \newcommand{\jetcasVC}[1]{#1}
 
 \else
 \newcommand{\TBC}[1]{\textcolor{blue}{\textbf{#1}}}
   \newcommand{\jetcasVB}[1]{\textcolor{blue}{\textbf{#1}}}
   \newcommand{\jetcasVC}[1]{\textcolor{blue}{\textbf{#1}}}
    
  \newcommand{\markblue}[1]{\textcolor{blue}{\textbf{#1}}}
  \newcommand{\todoh}[2]{}

 \fi
  
\newcommand{\unit}[1]{{[}#1{]}}
\newcommand{\tpu}[0]{Tile-PU}
\newcommand{\tpus}[0]{Tile-PUs}

\newcommand{\figref}[1]{Fig.~\ref{#1}}
\newcommand{\tabref}[1]{Tbl.~\ref{#1}}

\newcommand{\secref}[1]{Sec.~\ref{#1}}

\ifdefined\icml
\icmltitlerunning{Hyperdrive}
\fi

\begin{document}
\bstctlcite{IEEEexample:BSTcontrol}
\normalem

\ifdefined\icml
\twocolumn[

\begin{icmlauthorlist}
\icmlauthor{Renzo Andri}{ethiis}
\icmlauthor{Lukas Cavigelli}{ethiis}
\icmlauthor{Davide Rossi}{unibo}
\icmlauthor{Luca Benini}{ethiis,unibo}
\end{icmlauthorlist}

\icmlaffiliation{ethiis}{Integrated Systems Laboratory, ETH Zurich, Zurich, Switzerland}
\icmlaffiliation{unibo}{Uni Bo}
\icmlaffiliation{dummy}{Affiliation Hidden For Blind Review}

\icmlcorrespondingauthor{Renzo Andri}{renzo.andri@iis.ee.ethz.ch}

\icmlkeywords{Machine Learning, ICML}
\vskip 0.3in
]

\printAffiliationsAndNotice{}  
\fi

\title{Hyperdrive: A Multi-Chip Systolically Scalable Binary-Weight CNN Inference Engine}

\author{\IEEEauthorblockN
{Renzo Andri\IEEEauthorrefmark{1}, Lukas Cavigelli\IEEEauthorrefmark{1}, Davide Rossi\IEEEauthorrefmark{2}, Luca Benini\IEEEauthorrefmark{1}\IEEEauthorrefmark{2}}\\
\IEEEauthorblockA{\IEEEauthorrefmark{1}Integrated Systems Laboratory, ETH Zurich, Zurich, Switzerland \IEEEauthorrefmark{2}DEI, University of Bologna, Bologna, Italy} 
}

\maketitle

\newacronym{TPU}{TPU}{Tile Processing Unit}

\begin{abstract}
Deep neural networks have achieved impressive results in computer vision and machine learning. Unfortunately, state-of-the-art networks are extremely compute and memory intensive which makes them unsuitable for mW-devices such as IoT end-nodes. Aggressive quantization of these networks dramatically reduces the computation and memory footprint. Binary-weight neural networks (BWNs) follow this trend, pushing weight quantization to the limit. Hardware accelerators for BWNs presented up to now have focused on core efficiency, disregarding I/O bandwidth and system-level efficiency that are crucial for deployment of accelerators in ultra-low power devices. We present Hyperdrive: a BWN accelerator dramatically reducing the I/O bandwidth exploiting a novel binary-weight streaming approach, which can be used for arbitrarily sized convolutional neural network architecture and input resolution by exploiting the natural scalability of the compute units both at chip-level and system-level by arranging Hyperdrive chips systolically in a 2D mesh while processing the entire feature map together in parallel. Hyperdrive achieves \revONE{4.3}~TOp/s/W system-level efficiency (i.e., including I/Os)---\revONE{3.1}\x{} higher than state-of-the-art BWN accelerators, even if its core uses resource-intensive FP16 arithmetic for increased robustness.
\end{abstract}
\newcolumntype{H}{>{\iffalse}c<{\fi}@{}}


\section{Introduction}

Over the last few years, deep neural networks (DNNs) have revolutionized computer vision and data analytics. Particularly in computer vision, they have become the leading approach for the majority of tasks with rapidly growing data set sizes and problem complexity, achieving beyond-human accuracy in tasks like image classification. What started with image recognition for handwritten digits has moved to data sets with millions of images and 1000s of classes \cite{Russakovsky2014,He2015}. What used to be image recognition on small images \cite{LeCun2010,Sermanet2012} has evolved to object segmentation and detection \cite{Wu2016a,Ren2015,Long2015,Cavigelli2016a,Cavigelli2016b} in high-resolution frames---and the next step, video analysis, is already starting to gain traction \cite{Feichtenhofer2016,Scheidegger2017ImpactClassification,Cavigelli2017}. Many applications from automated surveillance to personalized interactive advertising and augmented reality have real-time constraints, such that the required computation can only be run on powerful GPU servers and data center accelerators such as Google's TPUs \cite{Jouppi2017}.

At the same time, we observe the trend towards ``internet of things'' (IoT), where connected sensor nodes are becoming ubiquitous in our lives in the form of fitness trackers, smart phones and surveillance cameras \cite{Conti2017,GapNews2018}. This creates a data deluge that is never analyzed and raises privacy concerns when collected at a central site \cite{Baraniuk2011}. Gathering all this data is largely unfeasible as the cost of communication is very high in terms of network infrastructure, but also reliability, latency and ultimately available energy in mobile devices \cite{Schulz2017}. The centralized analysis in the cloud also does not solve the compute problem, it merely shifts it around, and service providers might not be willing to carry the processing cost while customers do not want to share their privacy-sensitive data \cite{Merritt2018}. 

A viable approach to address these issues is edge computing---analyzing the vast amount of data close to the sensor and transmitting only condensed highly informative data \cite{Conti2017,Rusci2017}. This information is often many orders of magnitude smaller in size, e.g., a class ID instead of an image, or even only an alert every few days instead of a continuous video stream. However, this implies that the data analysis has to fit within the power constraints of IoT nodes which are often small-form factor devices with batteries of a limited capacity, or even devices deployed using a set-and-forget strategy with on-board energy harvesting (solar, thermal, kinetic, \dots)~\cite{Weddell2013}. 

Recently, several methods to train neural networks to withstand extreme quantization have been proposed, yielding the notions of binary- and ternary-weight networks (BWNs, TWNs) and binarized neural networks (BNNs) \cite{AojunZhou2016,Venkatesh2017,Courbariaux2015a}. BWNs and TWNs allow a massive reduction of the data volume to store the network and have been applied to recent and high-complexity networks with an almost negligible loss. 
In parallel, the VLSI research community has been developing specialized hardware architectures focusing on data re-use with limited resources and optimizing arithmetic precision, exploiting weight and feature map (FM) sparsity, and performing on-the-fly data compression to ultimately maximize energy efficiency \cite{Sze2017,cavigelli2018extended}. However, these implementations fall into one of two categories: 
1) They stream the entire or even partial FMs into and out of the accelerator ending up in a regime where I/O energy is far in excess of the energy spent on computation, hitting an energy efficiency wall: the state-of-the-art accelerator presented in \cite{Andri2016a} has a core energy efficiency of 59\,TOp/s/W, but including I/O power it is limited to 1\,TOp/s/W; or
2) they assume to store the entire network's weights and intermediate FMs on-chip. This severely constrains the DNN's size that can be handled efficiently by a small low-cost IoT-end node class chip.
It also prevents the analysis of high-resolution images, thus precluding many relevant applications such as object detection.

The main contributions of this work are:
\begin{enumerate}
    \item A new and highly optimized yet flexible core architecture systolically scalable to high-resolution images to enable applications such as object detection.
    \item \revONE{A new computational model, which exploits the reduced size of the weights due to the binarization in BWNs. As the size of the weights becomes much smaller than the intermediate feature maps, Hyperdrive streams the weights instead of the intermediate feature maps. With this new method, Hyperdrive enables execution of state-of-the-art BWNs on tiny, power-constrained chips, while overcoming the I/O energy-induced efficiency wall.} 
    \item An in-depth analysis of this architecture in terms of memory requirements, I/O bandwidth, and scalability including measurements of the chip implemented in GF 22\,nm~FDX technology, showing a \markblue{1.8\x{}} and \markblue{3.1\x{}} gain in energy efficiency in image classification and object detection, respectively, even though our core uses resource-intensive FP16 arithmetic for increased robustness.
    \item We show that the system is systolically scalable to multiple chips with the elementary chip size fixed to a maximum area constraint arranged in a 2D mesh operating on tiles of the entire feature map. The extension is also implemented in GF 22\,nm FDX technology and is evaluated on layout simulations, showing that even with the overhead of exchanging the border pixels, the I/O \revONE{energy} can be reduced up to \markblue{5.3\x{}} compared with state-of-the-art accelerators.
\end{enumerate}

The remainder of the paper is organized as follows. \secref{sec:related} presents a review of the previous works more closely related to the architecture presented in this paper. \secref{sec:systemarchitecture} and \secref{sec:computationalmodel} introduce the Hyperdrive architecture and computational model, respectively, mainly focusing on its key innovation aspect: stationary feature-map and streaming binary-weights for reduced I/O bandwidth and improved system-level energy efficiency. \secref{sec:systolicdesign} describes the extensions to the presented architecture enabling a systolic-scalable system composed of Hyperdrive chips. \secref{sec:results} presents the results of the chip implemented in 22nm FDX technology, providing details about its characterization, benchmarking, and comparison with respect to the state-of-the-art of binary-weight CNN accelerators. Finally, \secref{sec:conclusion} closes the paper with some final remarks.

\section{Related Work}
\label{sec:related}
\subsection{Software-Programmable Platforms}
The availability of affordable computing power on GPUs and large data sets have sparked the deep learning revolution, starting when AlexNet obtained a landslide victory in the ILSVRC image recognition challenge in 2012~\cite{Krizhevsky2012a}. Since then we have seen optimized implementations \cite{Chetlur2014,Cavigelli2015a,Cavigelli2016} and algorithmic advances such as FFT-based and Winograd convolutions further raising the throughput \cite{Vasilache2014,Lavin2015a}. The availability of easy-to-use deep learning frameworks (TensorFlow, Torch, Caffe, \dots) exploiting the power of GPUs transparently to the user has resulted in wide-spread use of GPU computing. With the growing market size, improved hardware has become available as well: Nvidia has introduced a product line of systems-on-chip for embedded applications where ARM cores have been co-integrated with small GPUs for a power range of 5-20\,W and $\approx$50\,GOp/s/W. 
Also, the GPU's architecture has been optimized for DNN workload, introducing tensor cores and 
fast half-precision floating-point (FP16) support. The latest device, Nvidia's V100, achieves 112\,TFLOPS at 250\,W \cite{nvidia2017}---an energy efficiency of 448\,GOp/s/W. Its best known competitor, the first version of Google's TPU \cite{Jouppi2017}, works with 8-bit arithmetic and achieves 92\,TOp/s at 384\,W (240\,GOp/s/W). With these power budgets, however, they are many orders of magnitude away from the power budget of IoT. \jetcas{Furthermore, they cannot exploit the advantages of many recent techniques to co-design DNN models for efficient computation.}

\subsection{Co-Design of DNN Models and Hardware}
\label{sec:optNet}
Over the last few years, several approaches adapting DNNs to reduce the computational demand have been presented. One main direction was the reduction of the number of operations and model size. Specifically, the introduction of sparsity provides an opportunity to skip some operations. By pruning the weights a high sparsity can be achieved particularly for the fully-connected layers found at the end of many networks and the ReLU activations in most DNN models injects sparsity into the FMs, which can be exploited \cite{Yang2017,Han2016a}.

A different direction is the research into reduced precision computation. Standard fixed-point approaches work down to 10-16\,bit number formats for many networks. It is possible to further reduce the precision to 8\,bit with small accuracy losses ($<1\%$) when retraining the network to adapt to this quantization \cite{Gysel2016a}. There are limitations to this: 1) for deeper networks higher accuracy losses (2-3\% for GoogLeNet) remain, and 2) Typically, only the inputs to the convolutions are quantized in this format. Internal computations are performed at full precision, which implies that the internal precision is very high for large networks, e.g., for a 3\x3 convolution layer with 512 input FMs, this adds 12\,bits. 
Further approaches include non-linearly spaced quantization in the form of mini-floats \cite{Gysel2016a}, and power-of-two quantization levels replacing multiplications with bit-shift operations (i.e., INQ \cite{AojunZhou2016}).

Several efforts have taken the path to extreme quantization to binary (+1/-1) or ternary (+1/0/-1) quantization of the weights while computing the FMs using floats. This massively compresses the data volume of the weights and has even been shown to be applicable to deep networks with an accuracy loss of approximately 1.6\% for ResNet-18 \cite{AojunZhou2016} and thus less than the fixed point-and-retrain strategies. The next extreme approach are (fully) binary neural networks (BNNs), where the weights and FMs are binarized \cite{Courbariaux2016}. While this approach is attractive for extreme resource constrained devices \cite{Rusci2017,Bahou2018}, the associated accuracy loss of 16\% on ResNet-18 is unacceptable for many applications.

\subsection{FPGA and ASIC Accelerators}
Many hardware architectures targeting DNNs have been published over the last few years. The peak compute energy efficiency for fixed-point CNN accelerators with precision bigger than 8\,bit can be found at around 50\,GOp/s/W for FPGAs, 2\,TOp/s/W in 65\,nm and around 10\,TOp/s/W projected to 28\,nm \cite{Cavigelli2016,Chen2016,Du2015,Conti}.

\jetcas{Many of the sparsity-based optimizations mentioned in \secref{sec:optNet} have been implemented in hardware accelerators \cite{Han2016a,Aimar2017} and achieve an up to 3\x{} higher core energy efficiency and raise the device-level energy efficiency by around 70\% through data compression. The effect of training DNNs to become BWNs simplifies the computations significantly and has shown the biggest impact on core compute-only energy with an energy efficiency of 60\,TOp/s/W in 65\,nm \cite{Andri2016a}. }

\jetcas{State-of-the-art silicon prototypes such as QUEST \cite{QUEST} or UNPU \cite{UNPU} are exploiting such strong quantization and voltage scaling and have been able to measure such high energy efficiency with their devices. The UNPU reaches an energy efficiency of 50.6\,TOp/s/W at a throughput of 184\,GOp/s with 1-bit weights and 16-bit activations on 16\,mm\textsuperscript{2} of silicon in 65\,nm technology. 
However, all the aforementioned implementations, either don't consider the necessary I/O energy for streaming the FMs in the computation of energy efficiency, or they assume that intermediate results can be entirely stored in the limited-size on-chip memory. This restricts these devices to run networks capable of solving only low-complexity image recognition tasks or re-introduces a system-level energy efficiency wall at around 1\,TOp/s/W as soon as the feature maps need to be streamed off-chip \cite{Andri2016a}.}

\jetcas{QUEST \cite{QUEST} addresses this issue by 3D-stacking 96\,MB of SRAM distributed across 8 dies using inductive coupling for die-to-die wireless communication. They apply 4-bit logarithmic quantization to both the weights and the feature maps, which results in an accuracy drop in excess of what BWNs with high-precision feature maps achieve. In this configuration, they obtain an energy efficiency of 594\,GOp/s/W while achieving a throughput of 1.96\,TOp/s at 3.3\,mW on a 122\,mm\textsuperscript{2} die in 40\,nm technology.
}

\jetcasVB{Hyperdrive not only exploits the advantages of reduced weight memory requirements and computational complexity, but fundamentally differs from previous BWN accelerators \cite{Andri2016a, Wang2017, UNPU}. The main concepts can be summarized as: 1) Feature Maps are stored entirely on-chip, instead the weights are streamed to the chip (i.e., feature map stationary). Thanks to the binary nature of the weights the overall I/O demand is reduced dramatically. 2) Through its hierarchically systolic-scalable structure it allows to efficiently scale to any sized feature map and even with silicon area restriction it is still scalable by tailing on a 2D mesh of Hyperdrive chips.}

\section{Hyperdrive Architecture}
\label{sec:systemarchitecture}

\begin{figure}
    \centering
    \includegraphics[width=0.90\columnwidth]{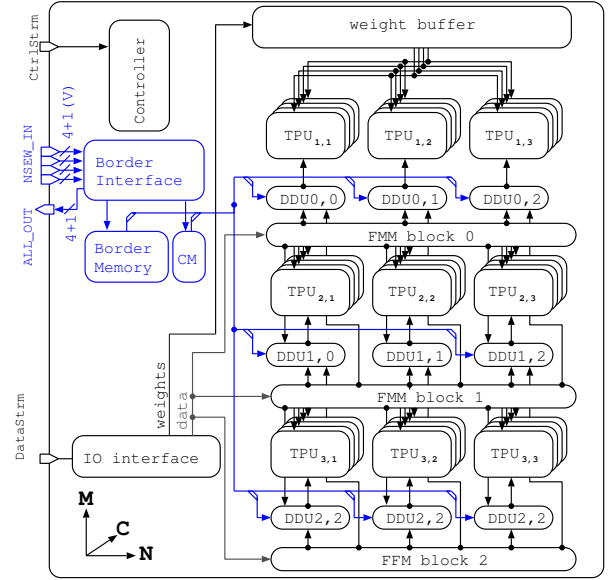}
    \caption{\revONE{System overview} with $C\times M\times N=4\times3\times 3$ tiles. Marked in blue are hardware block for the multi-chip systolic extension including the border interface which orchestrates any write and read to the border and corner memories and distributes it to the Data Distribution Units (DDUs). Furthermore, it sends and receives calculated pixels to and from the chip neighbors.}
    \label{fig:hyperdrive_top}
\end{figure}

\begin{figure}
    \centering
    \includegraphics[width=0.90\columnwidth]{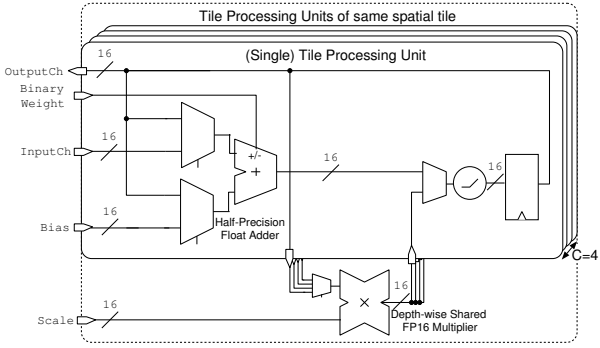}
    \caption{\revONE{\emph{Tile Processing Units (TPUs)} of same spatial tile (\mbox{\tpu$_{(\cdot, x, y)}$}): Every single \tpu{} (i.e., 4 shown in figure) provides a FP16 adder, accumulation register and ReLU activation unit. There is one time-shared FP16 multiplier per spatial tile and shared among the $C=4$ \tpus{} in the depth dimension, indicated by the dots. The FMs are calculated in a interleaved way for all $C$ output dimensions. The (single-bit) binary weight is applied as the sign input for the FP16 adder.}}
    \label{fig:tpu}
\end{figure}

\jetcasVC{Hyperdrive is a scalable and flexible binary-weight neural networks accelerator that can be parametrized to fit a wide range of networks targeting a variety of tasks from classification to object detection. \figref{fig:hyperdrive_top} shows a block diagram of Hyperdrive, where $M\times N$ indicate the spatial parallelism (i.e., size of the FM), while $C$ the output channel parallelism. It is composed of the following components:}
\begin{itemize}
\item \textit{Feature Map Memory (FMM)}: Is a multi-banked memory storing input and output FMs.
\item Array of $C\times M\times N$ \textit{Tile Processing Units (TPUs)}: A single \tpu{} is illustrated in \figref{fig:tpu}. It contains 1) a half-precision float adder/subtractor to accumulate partial sums of the output pixels, bias and the bypass input FM (in case of residual blocks), 2) a half-precision multiplier for the FM-wise batch-normalization shared among the Tile-PUs of the same tile, and 3) a ReLU activation unit. \TBC{Each Tile-PU$_{(c,x,y)}$ is operating on the spatial tile $(x,y)$ of the $M\times N$ tiles and on the output channel c from $C$}. Each Tile-PU is connected to its 8 spatial neighboring \tpus{} (i.e., directly adjacent \tpus) to quickly access neighboring pixels.
\item \textit{Weight Buffer (WBuf)}: Stores the weights of the current $C$ output FMs.
\item \textit{Data Distribution Units (DDUs)}: Distributes the data from the memories to the corresponding \tpu{} units or manages zero-padding.
\item \jetcas{\textit{Border and Corner Memory BM, CM}: Storage for pixels which are part of neighboring chips.}
\item \jetcas{\textit{Border Interface (BI/F)}: Sends and receive border pixels from/to neighboring chips and stores pixels into Border and Corner Memories.}
\end{itemize}

\begin{figure}
    \centering
    \includegraphics[width=0.75\columnwidth]{\figures{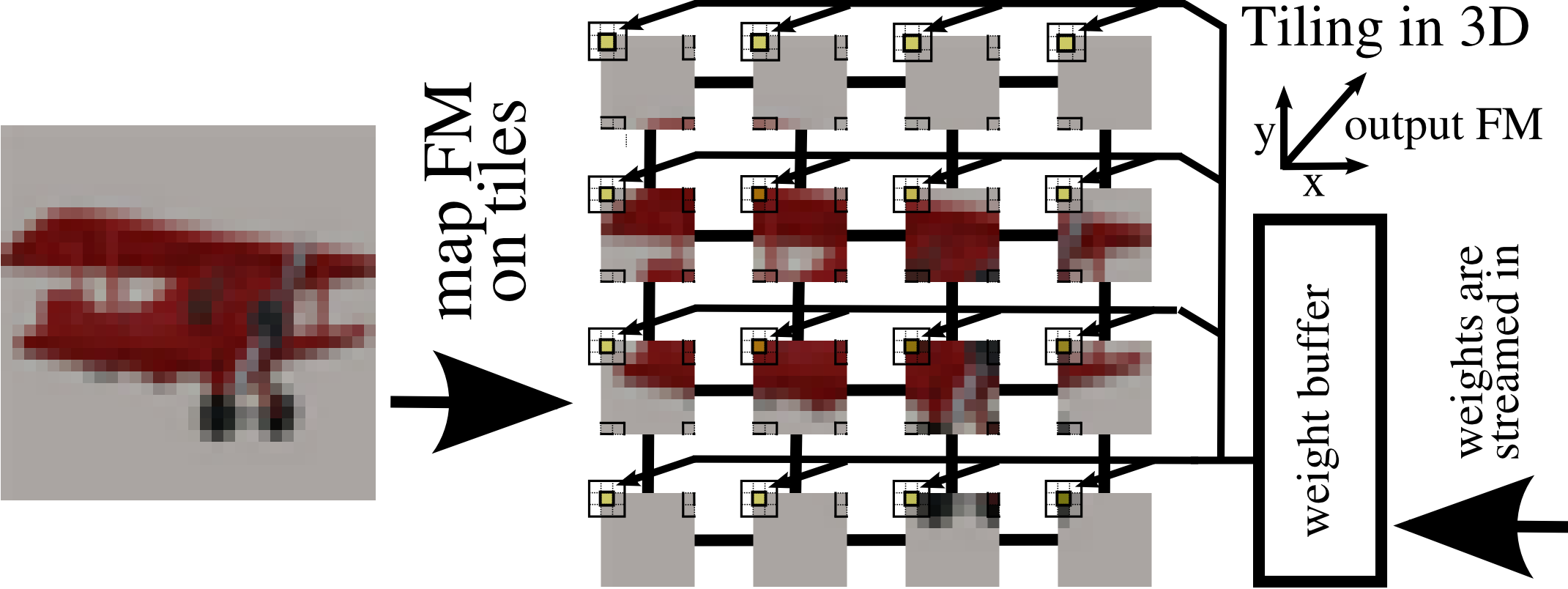}}\vspace{-0.1cm}
    \caption{The feature maps are tiled and processed in parallel \tpus{} \cite{Andri2018a}.}
    \label{fig:tiling}
\end{figure}

The superior efficiency of Hyperdrive is achieved exploiting data re-use at different levels:

\begin{itemize}
    \item Output FM level: \TBC{The output FMs are tiled into blocks of $C$ FMs which are calculated at the same time in the depth-wise parallel \tpus{} which allows to load the input FMs just once for $C$}
    \item Spatial level: The input FM is tiled into M\x N \revONE{equally-sized} image patches and calculated in parallel in the M\x N spatial processing units illustrated in \figref{fig:tiling}. Weights are read once from off-chip memory only and used to calculate all M\x N partial sums for the corresponding tiles. 
    \item Weight re-use: Weights are stored in the \textit{weight buffer}, which is implemented as a latch-based standard cell memory for optimal energy efficiency \cite{Andri2016a}.
    \item \jetcas{Border re-use: border pixels are transmitted only once to the corresponding neighbor chip and stored in its Border and Corner Memory instead of reading every time.}
\end{itemize}

\section{Computational Model}
\label{sec:computationalmodel}

\jetcas{State-of-the-art CNNs like ResNet-34 impose high demands in computational complexity and memory for the large space of parameters and intermediate Feature Maps. However, for BWNs, streaming the weights rather than the FMs or both is particularly attractive due to the compression by 16\x\ (i.e., from FP16).}

\TBC{CNNs are composed of several neural network layers, whereas the main building block are Convolution Layers which can be formulated as a mapping from the 3D input Feature Map space (i.e., $\text{FM}^\text{in}$) of $n_{in}$ channels with $h_{in}\times w_{in}$ sized spatial dimensions to the 3D output Feature Map space (i.e., $\text{FM}^\text{out}$) of $n_{out}\times h_{out}\times w_{out}$ size and can be described as follows:}
\begin{equation*} \label{eq:conv}
 \mathbb{R}^{n_{in}\times h_{in}\times w_{in}}\overset{CNN}{\rightarrow} \mathbb{R}^{n_{out}\times h_{out}\times w_{out}} 
\end{equation*}
\begin{equation*}
\text{FM}^\text{out}\mapsto \text{FM}^\text{in} \text{ s.t.}
\end{equation*}

\begin{equation*}
\text{FM}^\text{out}(c_{out}, \cdot, \cdot) =\beta_{c_{out}}+\alpha_{c_{out}}\hspace{-3mm}\sum_{c_{in}\in I_{n_i}}\hspace{-3mm}\text{FM}^\text{in}(c_{in},\cdot, \cdot)\ast k_{c_{out},c_{in}}(\cdot, \cdot)
\end{equation*}

\jetcasVB{Every single output channel $c_{out}$ is calculated by convolving all input feature maps $c_{in}$ with the corresponding filter kernel $k_{c_{out},c_{in}}\in\mathbb{R}^{h_k\times w_k}$, scaled by the factor $\alpha_{c_{out}}$ and accumulated to a bias term $\beta_{c_{out}}$. It should be noted here, that Batch normalization which are quite common after convolution layers, can be merged with biasing and scaling, as the coefficients stay constant after training.}

\subsection{\revONE{Principles of Operation}}\label{sec:princop}
The operations scheduling is summarized in Algorithm~\ref{alg:example} and illustrated in \tabref{tab:scheduling} for an implementation of the architecture featuring $C\times M\times N=16\times7\times7$ \tpu\ with 8\x 8 sized spatial tiles $\tilde{p}$ and for a 3\x3 convolution layer with 16\x64 FMs, whereas the output channels are tiled into blocks $\tilde{c}_{out}$ of $C=16$ channels. After the entire input feature map is loaded into the FMM, the system starts inferring the network. The output FM-level and spatial parallelism is indicated in lines 2 and 3, whereas every \tpu{} is working on its assigned spatial tile $\tilde{p}$ and output channel tile $\tilde{c}$ .

Then in the inner loop, the contribution for all pixels from the corresponding tile and output channel are calculated. From the streaming approach, a logical approach would be to load the weights and apply it to the entire patch for every \tpu, unfortunately, the patches can be large, and this introduces frequent writes and reads to random-access memory (FMM), instead the weights streamed to the chip are stored in a weight buffer (Line 11) which can be implemented in a small memory (i.e., latch-based memory for low energy) and where the weights for the current $C$ output channels (of all input channels) are stored. \revONE{In this way, we avoid writing and re-fetching intermediate FM values.}

The pixels are then calculated by iterating through all filter points (e.g., 9 in 3\x 3 kernels) and input channels $c_{in}$ (lines 7 and 8). On each cycle one binary weight per parallel feature map dimension $\#\tilde{c}_{out}$ is loaded from the weight buffer (Line 14) and input Feature Map pixel per spatial tile ($\#\tilde{p}=\#\{\text{\tpus}\}=M\cdot N$) are loaded from the FMM (Line 16). \revONE{All the \tpus{} access either their own FMM bank in case that the feature $p+\Delta$ (for the filter tap $\Delta$, e.g., (-1,-1) for the top-left weight of a 3\x3 filter) lies in the same tile $\tilde{p}$ or from the corresponding FMM bank of the corresponding neighboring \tpu. All these accesses are aligned (e.g., all the \tpus{} are reading the FMM bank of their corresponding top-left neighbor) and therefore no access conflicts occur}. The weights are multiplied with the binary weights: this is implemented as a change of sign and then accumulated with the previous partial sum $v$ (Line 17). When all contributions for all input channels and filter taps have been accumulated, a scaling factor (e.g., from batch normalization) is applied to it (Line 21), bypass is added (Line 22) and finally the channel bias is added (Line 23), before it is written back to the feature map memory (Line 24).

\TBC{Bypass paths are common in several CNNs like ResNet-34 and are shown in \figref{fig:basicresidual}. As will be explained in the next section, the bypass can be read, added to the partial sum and stored back to the same memory address avoiding additional memory for the bypass FM. Unfortunately, this does not work in the same cycle, therefore adding the bias (Line 21) has been moved after the bypass (Line 20) and stalling can be avoided.}

\begin{table*}[t]
\caption{Time schedule for a 16 input FM and 64 output FM 3\x3 convolution. Notation for filter weights: $f_{\text{input FM}, \text{output FM}}^{\text{filter tap} (\Delta y,\Delta x)}$.}\label{tab:scheduling}
\begin{adjustbox}{max width=\textwidth}
\begin{tabular}{|c|c|c|c|c|c|c|c|c|c|c|c|c|c|c|c|c|} \hline
cycle & 1& 2& ...& 16& 17& ...& ...& 144& 145& ...& 288& ...& 9216&9217& ...&36.8k \\ \hline
weight input & $f_{1, (1-16)}^{-1,-1}$& $f_{2, \cdot}^{-1,-1}$ & ... & $f_{16, \cdot}^{-1,-1}$ & $f_{1, \cdot}^{-1,0}$ &... &... & $f_{16, \cdot}^{1,1}$ & \multicolumn{5}{c|}{No I/O (loaded from weight buffer)} & $f_{1, (17-32)}^{-1,-1}$ &... & No I/O \\ \hline
input FM & 1 & 2 & ... & 16 & 1 & ... & ... & 16 & 1 & ... & 16 & ... & 16 & 1 & ... & 16 \\ \hline
filter tap pos. & \multicolumn{4}{c|}{ -1,-1 } & \multicolumn{2}{c|}{ -1,0 } & ... & +1,+1 & -1,-1 & ... & +1,+1 & ... & +1,+1 & -1,-1 & ... & +1,+1 \\ \hline
outp. pixel pos. & \multicolumn{8}{c|}{ 1,1 } & \multicolumn{3}{c|}{ 1,2 } & ... & 8,8 & 1,1 & ... & 8,8 \\ \hline
output FM & \multicolumn{13}{c|}{ 1-16 (in parallel) } & 17-32 & ... & 49-64 \\ \hline
\end{tabular} 
\end{adjustbox}
\end{table*}

\begin{algorithm}
  \caption{Hyperdrive Execution-Flow}
  \label{alg:example}
\begin{algorithmic}[1]
\REQUIRE{All input feature maps in $\text{FMM}^\text{in}$}
\REQUIRE{Weight Stream}

\FORALL{\revONE{$M\times N$ pixel tiles} $\tilde{p}$ (in parallel HW units)} 
    \FORALL{\revONE{$C$ output channel tiles} $\tilde{c}_{out}$ (in parallel HW units)}
        \STATE \textcolor{gray}{\tpu\ for output channel tile $\tilde{c}_{out}$ and pixel tile $\tilde{p}$}\;
        \STATE def readFMfromMemory
        \FORALL{output channel $c_{out}$ in tile $\tilde{c}_{out}$}
            \STATE{v = 0}
            \FORALL{pixel $p=(y,x)$ in tile $\tilde{p}$}
                \FORALL{filter points $\Delta = (\Delta y, \Delta x)$  with \newline\ \ $\Delta y = -\lfloor \frac{h_k}{2}\rfloor,...,-1,0,1,..., \lfloor \frac{h_k}{2}\rfloor, $\newline\ \ $\Delta x = -\lfloor \frac{w_k}{2}\rfloor,...,-1,0,1,..., \lfloor \frac{w_k}{2}\rfloor $}
                    \FORALL{input channel $c_{in}$}
                      \IF{$w[c_{in}, c_{out}, \Delta]$ $\not\in$ WBuf }
                        \STATE $k_{c_{out}, c_{in}}(\Delta) = \text{wghtStrm}$\;
                        \STATE{WBuf$[c_{in}, c_{out}, \Delta] = k_{c_{out}, c_{in}}(\Delta)$}
                      \ENDIF
                      \STATE $w=\text{WBuf}[c_{in}, c_{out}, \Delta]$ (read of \#$\tilde{c}_{out}$ bit)\;
                      \STATE \textit{\revONEnotbold{// Aligned read of $\text{FMM}^{in}[p+\Delta,c_{in}]$ from // corresponding memory bank (either from // its own memory bank or the correspond- // ing neighbor's bank).}}
                      \STATE $x= \text{FMM}^{in}[p+\Delta,c_{in}]$ (read of \#$\tilde{p}$ words)\;
                      \STATE $v = (v+x \cdot w)=\begin{cases}
                                   v+x  & \text{if } w=1 \\
                                   v-x  & \text{otherwise}
                                  \end{cases}$\;
                    \ENDFOR
                \ENDFOR
            \ENDFOR
            \STATE (opt) $v \mathrel{{*}{=}} \text{bnorm}(c_{out})$\;
            \STATE (opt) $v \mathrel{{+}{=}} \text{FMM}^\text{bypass}(c_{out}, p)$\;
            \STATE (opt) $v \mathrel{{+}{=}} \text{bias}(c_{out})$\;
            \STATE $\text{FMM}^\text{out}[c_{out}, p] = v$ (save in memory)\;
        \ENDFOR

    \ENDFOR
\ENDFOR

\end{algorithmic}
\end{algorithm}

\subsection{CNN Mapping}
\label{sec:evalresnetmem}
\begin{figure*}
    \centering
    \includegraphics[width=0.75\linewidth]{\figures{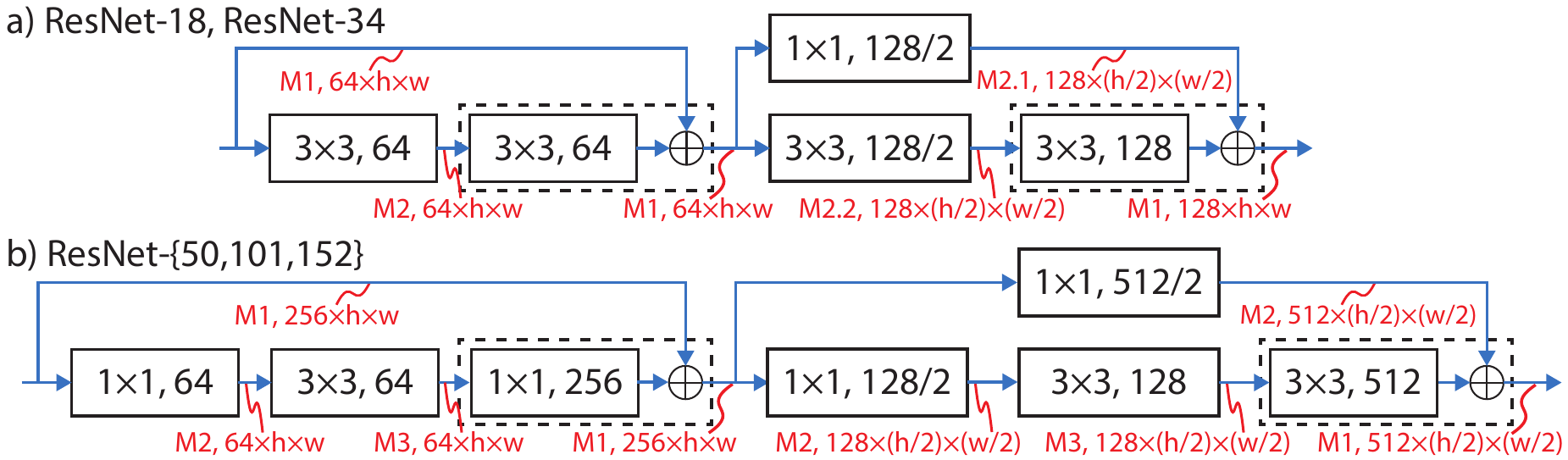}}
    \caption{Early block of layers of ResNet-34 and transition to next type of layer block. Activation and batch normalization layers are not indicated separately. Dashed rectangles imply on-the-fly addition to eliminate the need for additional memory}
    \label{fig:basicresidual}
\end{figure*}

The size of the on-chip memory for intermediate FM storage has to be selected depending on the convolution layer with the largest memory footprint of the network, hereinafter referred as \textit{Worst-Case Layer} (WCL). Typically, the WCL is at the beginning of the network, since a common design pattern is to double the number of FMs after a few layers while performing at the same time a 2\x2 strided operation, thereby reducing the number of pixels by 4\x\ and the total FM volume by 2\x. To perform the computations layer-by-layer, avoiding usage of power hungry dual-port memories, we leverage a ping-pong buffer mechanism reading from one memory bank and writing the results to a different memory bank. Hence, for a generic CNN the amount of memory required by the WCL is: $\max_{\text{layers in CNN}} n_\text{in}h_\text{in}w_\text{in}+n_\text{out}h_\text{out}w_\text{out}$ words, since all input and output FMs have to be stored to implement the described ping-pong buffering mechanism.

However, many networks have bypass paths, hence additional intermediate FMs have to be stored, as described in {\figref{fig:basicresidual}a} for the potential WCLs of ResNet-34. This aspect has two implications: \\
1) In order to avoid additional memory (+50\%), we perform an on-the-fly addition of the bypass path after the second 3\x3 convolution (i.e., the dashed rectangle is a single operation). This is done by performing a read-add-write operation on the target memory locations. \\
2) To avoid adding a stall cycle when reading and writing to the same memory area within the same cycle, the bias adding is moved after the bypass such that the following order is followed convolution, scale, bypass, bias, store back. In this way, the data can be read from memory address and stored back to the same address with one cycle latency.
3) The common transition pattern with the 2\x2-strided convolution does not require additional memory. It temporarily needs three memory segments, but two of them are 2\x{} smaller and can fit into what has been a single memory segment before (\texttt{M2} is split into two equal-size segments \texttt{M2.1} and \texttt{M2.2}).

\jetcasVB{In the following section, the calculation of the WCL for ResNet-like networks with basic bypass blocks is discussed in detail and numbers are presented for ResNet-34, but does not limit the execution of networks with smaller WCL.} To reduce off-chip data communication to a minimum, we will split the WCL in memory segments \texttt{M1, M2, \dots} to indicate which data needs to be kept in on-chip memories at the same time. Hyperdrive always operates on a single convolutional layer at a time and is iterating several times over the entire input FM which therefore needs to be stored on-chip in memory section \texttt{M1}. The same is valid for the output FM which is calculated and stored in \texttt{M2}, respectively. 

There are $n_{out}$ output channels which have a $h_{out}\times w_{out}$ sized output FM. These output FMs are calculated as sum of convolutions of every $n_{in}$ input channel (with FMs size of $h_{in}\times w_{in}$) on the $h_k\times w_k$ sized filter kernels $w_{k,n}$.

For a normal convolution layer, 
\begin{equation*}
\texttt{M}=\texttt{M1}+\texttt{M2}=n_{in}\cdot h_{in}\cdot w_{in}+n_{out}\cdot h_{out}\cdot w_{out} \text{ [words]}
\end{equation*}
need to be stored, because the entire input FM is needed to calculate every single output FM.

In a next step, the special case with residual bypasses is evaluated like in ResNet \cite{He2015} and similar residual networks. ResNet has two different types of residual blocks: the basic building block and the bottleneck building block. The basic building block is presented in \figref{fig:basicresidual}.

Within the basic building block, there are two different cases, the first is $n_{in}=n_{out}$ where there is no striding, thus also $h_{in}=h_{out}$ and $w_{in}=w_{out}$. The input FM to the residual block will then be placed in the virtual memory section \texttt{M1} and Hyperdrive computes the first $3\times 3$ convolution layer and writes the results into section \texttt{M2}, secondly Hyperdrive calculates the second convolutions layer reading from \texttt{M2} and accumulating the output FM with the bypassed values in \texttt{M1} on-the-fly and writing them back to \texttt{M1}. A total amount of 401\,kwords need to be stored.
\begin{align*}
\texttt{M}&=\texttt{M1}+\texttt{M2}& = & \ 2\cdot \texttt{M1}&=&\ 2n_{in}\cdot h_{in}\cdot w_{in} \\
  \texttt{M1}&=\texttt{M2}& = &n_{in}\cdot h_{in}\cdot w_{in}       
\end{align*}
\begin{align*}
M_{max}=2n_{in}\cdot h_{in}\cdot w_{in}=2\cdot64\cdot 56\cdot 56=401k \text{words} 
\end{align*}

In case of down-sampling the number of output channels is doubled $n_{out}=2n_{in}$ and the image sizes are reduced by 4\x{} to $h_{out}\times w_{out}=\frac{1}{2}h_{out}\times \frac{1}{2}w_{out}$. Also, the bypass needs to be strided. He \emph{et al.} suggest to either use the strided identity or to perform $1\times 1$ strided convolution, we will consider this case as it is more memory critic\revONE{al} than with subsampling \cite{He2015}. The input FM is read from \texttt{M1} and the $3\times 3$ strided convolution is performed and saved in \texttt{M2}, then the $1\times 1$ strided convolution on the bypass is evaluated and saved in \texttt{M3}, finally the 2nd convolution layer is performed on the data in \texttt{M2} and accumulated to the strided bypass in and to \texttt{M3}. It can be shown, that \texttt{M2} and \texttt{M3} are a quarter of the size of \texttt{M1} and 301\,kwords are needed for the three memory sections.
\begin{align*}
\texttt{M}&=&\texttt{M1}+\texttt{M2}+\texttt{M3}&=&1.5\cdot\texttt{M1}\\
&=&\ 1.5n_{in}\cdot h_{in}\cdot w_{in} \\
\texttt{M1} & = & \ n_{in}\cdot h_{in}\cdot w_{in}\\
\texttt{M2}=\texttt{M3}& = & \ 2n_{in}\cdot 0.5\cdot h_{in}\cdot 0.5\cdot w_{in}&=&0.5\cdot\texttt{M1} \\
\end{align*}
Due to the reduced size of the FM after every subsampling, just the first residual block need to be considered for dimensioning the memories. For ResNet-18 and ResNet-34, this translates to 401\,kwords which are 6.4\,Mbit with FP16.

Deeper residual networks (e.g., ResNet-50) are composed of the bottleneck building block (illustrated in \figref{fig:basicresidual}b), to evaluate the WCL, there are two cases to consider: with and without subsampling. In the first case, the input FM is stored in \texttt{M1} and needs to be stored for the entire bottleneck block. The output FM for the first 1\x 1 convolution layer is stored in \texttt{M2} and is 4\x{} smaller due to the 4\x{} smaller number of channels, then the 3\x 3 convolution layer calculates its features from \texttt{M2} to \texttt{M3} and the second 1\x1 convolution layer is calculated on-the-fly adding to the bypass FM.
\begin{align*}
\texttt{M}&=&\texttt{M1}+\texttt{M2}+\texttt{M3}&=&1.5\cdot\texttt{M1}\\
&=&\ 1.5n_{in}\cdot h_{in}\cdot w_{in} \\\\
\texttt{M1} & = & \ n_{in}\cdot h_{in}\cdot w_{in}\\
\texttt{M2}=\texttt{M3}& = & \ \frac{n_{in}}{4} \cdot h_{in} \cdot w_{in}&=&0.5\cdot\texttt{M1} \\
\end{align*}
In total 1.5\x{} of the input FM size is needed to evaluate the bottleneck block without subsampling. 
In case with subsampling, already after the 1\x1 convolution, the bypass needs to be evaluated which is another 1\x 1 convolution which we can map into \texttt{M4} memory. Instead of writing the feature map for the 3\x 3 convolution to \texttt{M3}, it can be written to \texttt{M1}, because this data is not needed any more. The 2\textsuperscript{nd} 1\x 1 convolution is then calculated on the fly from \texttt{M1} and \texttt{M4} back to \texttt{M1}.

\begin{align*}
\texttt{M}&=&\texttt{M1}+\texttt{M2}+\texttt{M4}&=&1.675\cdot\texttt{M1}\\
&=&\ \frac{13}{8}n_{in}\cdot h_{in}\cdot w_{in} &=& 1.2 M\text{words} \\\\
\texttt{M1} & = &\omit\rlap{$\max\left(n_{in}\cdot h_{in}\cdot w_{in}, \frac{2n_{in}}{4}\cdot \frac{h_{in}}{2}\cdot \frac{w_{in}}{2}\right)$}\\
&=&\ n_{in}\cdot h_{in}\cdot w_{in}  \\\\
\texttt{M2}& = & \ \frac{2n_{in}}{4} \cdot \frac{h_{in}}{2} \cdot \frac{w_{in}}{2}&=&0.125\cdot\texttt{M1} \\
\texttt{M4}& = & \ 2n_{in} \cdot \frac{h_{in}}{2} \cdot \frac{w_{in}}{2}&=&0.5\cdot\texttt{M1} \\
\end{align*}

This leads to a WCL of 1.2\,Mword or 19.2\,Mbit (Conv2) for ResNet-50/-152/\dots independently of the depth which would be 6.3\,mm${}^2$ of SRAM (0.3\,\textmu m${}^2$/bit in GF\,22nm FDX).

\begin{table}
\centering
\caption{Data Comparison for various typical networks with binary-weights and 16-bit FMs for single-chip implementation considering single-chip implementation (Top: Image Recognition, Bottom: Object Detection)}
\begin{tabular}{lcrrr}
\toprule
     network                & resolution & \makecell{weights\\\unit{bit}} & \makecell{all FMs\\\unit{bit}} & \makecell{WC mem.\\\unit{bit}}\hspace{-0.2cm} \\ 
     \midrule 
     ResNet-18\hspace{-0.0cm} & 224\x 224 & 11M       & 36M     & 6.4M   \\ 
     ResNet-34\hspace{-0.0cm} & 224\x 224 & 21M       & 61M     & 6.4M   \\ 
     ResNet-50\hspace{-0.0cm} & 224\x 224 & 21M       & 156M     & 21M \\ 
     ResNet-152\hspace{-0.0cm}& 224\x 224 & 55M      & 355M     & 21M \\ \midrule
     ResNet-34\hspace{-0.0cm} & \hspace{-0.3cm}2048\x 1024\hspace{-0.3cm} & 21M     & \calc{1}{2.5}G      & 267M\\ 
     ResNet-152\hspace{-0.0cm} & \hspace{-0.3cm}2048\x 1024\hspace{-0.3cm} & 55M     & \calc{1}{14.8}G      & 878M\\ 
       \bottomrule
\end{tabular}\\
\label{tab:datacomparison}
\end{table}

\subsection{\jetcasVB{Supported Neural Network Topologies}}
\jetcasVB{In the previous section, we have discussed the requirements to map the different ResNet-style networks onto Hyperdrive. For its implementation, we have parameterized the architecture to fit the feature maps of ResNet-34 on-chip. Nevertheless, Hyperdrive is neither restricted to these networks nor these applications---in fact, its scalability to multiple chips to process high-resolution images for object detection and image segmentation is a key feature of its architecture. For example, running the feature extraction for object detection using YOLOv2 \cite{Redmon2017} is supported by Hyperdrive. For the worst-case layer in terms of memory when processing $448\times 448$ pixel frames, we would need to be able to store 3.2\,M words---scaling up the memory by $2\times$ over the ResNet-34 parameterization would be sufficient to even run it even on a single chip, and for higher resolutions the workload and memory for the feature maps could be split across multiple chips as described in \secref{sec:borderextension}. Also, the Fire module of the size-optimized SqueezeNet \cite{Iandola2016} and SqueezeDet \cite{Wu2016a} topologies is supported by Hyperdrive. The grouped convolutions and shuffling operations present in MobileNetV2 \cite{sandler2018mobilenetv2} and ShuffleNet \cite{Zhang2017} can also be applied with the presented architecture. Also the not very common depth-wise separable convolutions present in some layers of MobileNetV2 can be computed using Hyperdrive, although not at maximum performance due to limited bandwidth of the on-chip SRAMs (no local re-use of the input feature map data possible). }

\jetcasVB{The only limitation is that several networks feature a first convolution layer with an exceptionally large kernel size (e.g., $7\times7$ convolution for both ResNet and YOLOv2, but making up less than 2\% of all operations). As Hyperdrive supports only $1\times1$ and $3\times3$ convolution layers, this first layer has to be computed off-chip before loading the data into Hyperdrive, or a small dedicated on-chip accelerator for the first layer could be included, which would perform these operations as the feature maps are streamed into the device. Networks optimized for compute effort, such as TinyYOLO \cite{Redmon2016} or MobileNetV2 \cite{sandler2018mobilenetv2}, are often only composed of $3\times3$ and $1\times1$ convolution layers and do not have such a first filter with an exceptionally large kernel size.}

\section{Scalability to Multiple Chips}\label{sec:borderextension}
\label{sec:systolicsystemscaling}\label{sec:systolicdesign}
\jetcasVB{Even though, we could show that the architecture is in theory scalable to any sized networks, the WCL is setting a real-world limit to it. Already ResNets with bottleneck layer require 19.2\,Mbit\footnote{Note that the WCL for ResNet-like networks does not depend on depth, but on the size of the images (e.g., 224\x224) and the building blocks (basic bypass in \figref{fig:basicresidual}a or bottleneck in \figref{fig:basicresidual}b). See also \tabref{tab:datacomparison} for a comparison of the WCLs.} to perform inference on small 224\x224 sized images and larger images (e.g., in typical object detection tasks) need 10s or 100s of Mbit. This clearly exceeds the area constraints of few Mbit in low-cost chip fabrication due to high production costs and diminished production yield.} A very natural solution is to extend the systolic architecture to multiple chips, in this way the feature map is first tiled on an array of $m\times n$ Hyperdrive chips and further tiled within each chip on their $M\times N$ Tile Processing Units, such that $M\cdot m\times N\cdot n$ tiles are operated in parallel.
\begin{figure}[H]
    \centering
    \includegraphics[width=0.6\columnwidth]{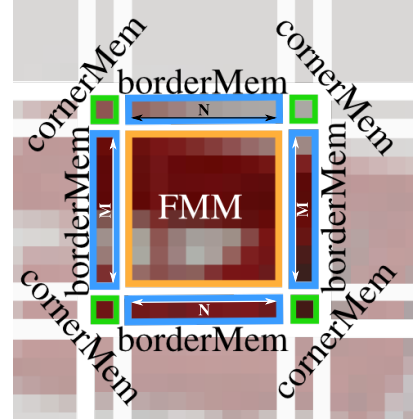}
    \caption{Memory Allocation in the multi-chip setup with 1\x 1 sized tiles for 3\x 3 sized kernels. The M\x N ``core'' tiles and pixels are stored in the FMM and the pixels located and calculated in the chip neighbor are stored in Border and Corner Memory. The Border Memory stores these M\x{} or N\x{} pixels (i.e., $7\times 16\,\text{bit}$) which can be accessed in the same cycle.}
    \label{fig:mem_allocation}
\end{figure}

\begin{figure*}
    \centering
    \includegraphics[width=\linewidth]{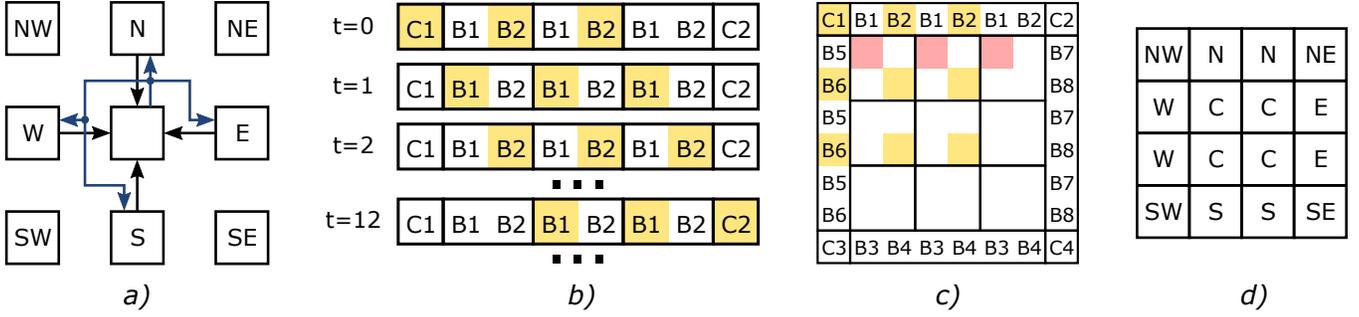}
    \caption{Multi-chip Considerations: \textbf{a)} Intra-chip connection: 1 output interface and 4 inputs from/to 4 direct neighbors, \textbf{b)} Border Memory and Corner memory access with address block ($c_{in}=1, h_k=w_k=3)$ for every single cycle \textbf{c)} Access pattern in case of a corner access: two reads from Border Memory (top and left) and one read from Corner Memory \textbf{d)} Chip Types in a systolic chip setting (North West to South East and Center chip)}
    \label{fig:memoryFigs}
\end{figure*}
\jetcas{Similarly to the single-chip setup, the Tile Processing Units need to access neighboring pixels, but in the multi-chip setup they might even lie on a different chip instead of just another tile memory. Three solutions are possible in this case, 1) the input feature maps of all chips are padded with the missing pixels, but this is not reasonable for deep networks as the padding increases steadily with the number of layers. 2) The border pixels are read from the neighboring chips when they are used, but this introduces high bandwidth requirement, as these pixels are needed several times or 3) the border pixels are sent once after they have been calculated to the neighboring chips and stored locally there. Hyperdrive implements option 3 which introduces two additional memories: A Border Memory BM and Corner Memory CM and have been added to the general architecture of Hyperdrive in \figref{fig:hyperdrive_top}.

\figref{fig:mem_allocation} illustrates the locations of the pixels from a chip perspective and \figref{fig:memoryFigs}a shows the perspective of a single chip connected to its neighboring chips which are overall arranged in a systolic way. Pixels residing in the border of the neighboring chips are stored in the Border Memory and pixels residing in the corners of the diagonal neighboring chips are stored in the Corner Memory and are read from there in case border pixels are requested by the computational model.
}
\subsection{Access Pattern and Storing Scheme of the Border Memories}
\figref{fig:memoryFigs}c illustrates the pixels and their memory location which are read in case of a corner pixel and \figref{fig:memoryFigs}b for all cases of access top border pixels. When border pixels but not corner pixels have to be accessed, one pixel per corresponding border \tpus{} is read and stored into the same memory block. In case of a corner, actually $M-1$ and $N-1$ pixels from two border sides (i.e., one vertical and one horizontal) and one corner pixel. Therefore, the border memory is split into two physically separated memory blocks allowing to read from both sides without the need of two-port memories or introducing any latency or stalls. Furthermore, chips are assigned a location chip type, which indicates which part of the feature map the chip is working on. They have been named corresponding to cardinal orientation: corner chips (NW, NE, SW, SE), border chips (N, W, E, S) and Center like illustrated in \figref{fig:memoryFigs}d. All chips sharing the same orientation work identically and synchronized, thus the exact position does not matter.\\

\subsection{Border and Corner Exchange}
Whenever a border pixel (e.g., N border) has been calculated, it is sent to the corresponding neighbor (i.e., south neighbor) and a flag is set indicating that it is waiting the same kind of pixel from its opposite neighbor (i.e., north neighbor). 

When a corner pixel (e.g., NW) is calculated, the pixel needs to be send to all three neighboring chips in the corresponding direction (N, W, NW). As the number of these pixels is small and to keep the inter-chip wiring small, no additional diagonal interfaces are introduced, but these pixels are forwarded by the corresponding vertical neighbor (N) to the diagonal chip (NW). Additionally, there are for every corner 2 additional flags which are set in the Border Interface: one for the forwarding chip (sending, N) and the receiving chip (NW). 
\subsection{Border and Corner Memory}
There are two different access patterns. If a corner pixel is accessed, the corner pixel, $N-1$ vertical pixels (left or right) and $M-1$ horizontal pixels (top or bottom) and one pixel need to be read from the corner memory, which is illustrated in \figref{fig:memoryFigs}c. In the other border cases, they are either $N$ vertical pixels or $M$ horizontal pixels (e.g., in \figref{fig:memoryFigs}b at $t\in \{1,2\}$). Therefore, the border memory can be seen as a horizontal or vertical extension to the FMM and $N$ and $M$ words can be read in a single cycle. As for the FMM, splitting the border memory into two physically separated memory blocks allows to read from both in the same cycle without introducing any additional latency. The memory needs to fit the overlapping border of the WCL whereas the width depends on the kernel width of the current and next layer. The overlapping rows or columns are $\lfloor \frac{h_k}{2}\rfloor $ or $\lfloor \frac{w_k}{2}\rfloor $ wide and can be determined directly from the WCL evaluation for FMM by dividing the spatial area and multiplying by the sum of all overlapping border rows or columns (which might differ for input and output FM). In case of ResNets with the basic building block (e.g., ResNet-34). The required memory for the left, right, top and bottom border (i.e., $\texttt{M}_{b,left}$, $\texttt{M}_{b,right}$, $\texttt{M}_{b,top}$, $\texttt{M}_{b,bottom}$) can therefore be calculated as follows:

\begin{align*}
\texttt{M}_{border}&=\texttt{M}_{b,left}+\texttt{M}_{b,right}+\texttt{M}_{b,top}+\texttt{M}_{b,bottom}
\\&=\texttt{M}\ \frac{2h_{in}+2w_{in}}{h_{in}\cdot w_{in}}=\texttt{M}\ \frac{2\cdot 56+2\cdot 56}{56\cdot56}=459\,\textnormal{kbit} 
\\ \texttt{M}_{b,left}&=\texttt{M}_{b,right}=2\left(n_{in} w_{in}\lfloor \frac{w_{k,l}}{2}\rfloor+n_{out} w_{out}\lfloor \frac{w_{k,l+1}}{2}\rfloor\right)
\\ \texttt{M}_{b,top}&=\texttt{M}_{b,bottom}=2\left(n_{in} h_{in}\lfloor \frac{h_{k,l}}{2}\rfloor+n_{out} h_{out}\lfloor \frac{h_{k,l+1}}{2}\rfloor\right)
\end{align*}
which is an increase of 7\% of overall memory.

The Border Memory (as indicated in \figref{fig:hyperdrive_top}) is then implemented with 4 high-density single-port SRAMs with 1024 lines of $7\cdot 16=112$.

The Corner Memory needs to store the diagonally overlapping pixels, which are $\floor*{\frac{h_k}{2}}\cdot\floor*{\frac{w_k}{2}}$ sized patches. In contrary to the discussions regarding the FMM and BM, the Corner Memory does not profit from striding such that for ResNet typed networks the last layer has the highest memory demand. Overall it can be dimensioned for ResNet-34 as
$(n_{in}+n_{out})\cdot4\floor*{\frac{h_k}{2}}\cdot\floor*{\frac{w_k}{2}}=2\cdot 512\cdot4\cdot 1\cdot 1\cdot 16\,\text{bit} = 64\,\text{kbit}$
which is another 1\% increase of overall memory. This memory has been implemented with a single-port memory of 4096 of 16-bit words.

\subsection{Interface Implementation}

During the computation of border pixels, every border \tpu{} sends and receive the pixels to/from the respective Border Interfaces. The border interfaces, placed on the 4 sides (as illustrated in \figref{fig:memoryFigs}a) of the accelerator, are responsible for buffering and transmitting pixels from/to the neighboring chips, synchronizing execution of the \tpus{} as well. For vertical and horizontal borders there is one $m\cdot C=7\cdot16=112$ entries buffer. \jetcasVB{When the buffer is non-empty, the border interface sends these pixels in an interleaved way and split into blocks of 4\,bits and 1 valid bit to the neighbors. Every chip itself has 4 in-coming serial interfaces from the directly adjacent neighbors (i.e., N, S, W, E). When data is received, it is de-serialized, recovered in its original 16-bit format and stored in the border/corner memories. The interfaces are also responsible for calculating the addresses of pixels received and transmitted from/to neighboring chips in the border memory. \figref{fig:tpu} shows in blue the extension needed for exchanging the borders between the chips with 1 out-going and 4 in-going intra-chip interfaces.}

\section{Experimental Results}
\label{sec:results}
The number of tiles has been chosen to be $M\times N = 7\times 7$, which allows for 4\x{} striding on 112\x 112 sized input FMs (like in common ResNet-like networks), while keeping all the TPUs busy with at least one single spatial pixel during the entire network. 
\revONE{We use the half-precision floating point (FP16) number format for the FMs as a conservative choice to ensure loss-less inference even for deeper networks \cite{migacz20178,das2018mixed}. Fixed-point or other alternative formats \cite{tagliavini2018transprecision} could be used to reduce the energy cost of the arithmetic operations. Fixed-point arithmetic units featuring a smaller number of bits (e.g., 8) would linearly impact the size of the on-chip memory for the FMs. By using FP16, the final accuracy is determined by the selected network and the corresponding BWN training algorithm. A ResNet-18 trained on the ImageNet dataset can run on Hyperdrive with a 87.1\% top-5 accuracy using the SBD-FQ training method \cite{hu2018training} (full-precision top-5 accuracy: 89.2\%).
}

The on-chip memory was sized to fit the WCL of ResNet-34 with 6.4\,Mbit (400\,kword) and is implemented with $M\times 8=7\times 8$ high-density single-port SRAMs with 1024 lines of $N\cdot 16=7\cdot 16=112$-bit words, whereas the memories are assigned to the $(M\times N)$ tiles. The output FM parallelism has been fixed to $C=16$. The weight buffer has been implemented to fit up to 512 (max. \#input FMs) $h_k\times w_k =3\times 3$ kernels for 16\x{} depth-wise parallelism. If more input FMs are needed, they can be tiled to 512 blocks and partial output FM can be calculated and summed up on-the-fly using the bypass mode. The frequently-accessed weight buffer has been implemented as a latch-based standard cell memory (SCM) composed of 5\x 8 blocks of 128 rows of 16-bit words, reducing the access energy to SRAM memories by 43\x{} \cite{Andri2016a}. \jetcasVB{It should be noted that even though the energy efficiency of SCMs are much better than SRAMs, they are also up to 8\x{} larger in area which limits this kind of memories to comparably small buffers (i.e., weight buffer), but not for the feature map memory.}

\subsection{Implementation Results}\label{sec:implRes}
\begin{figure}
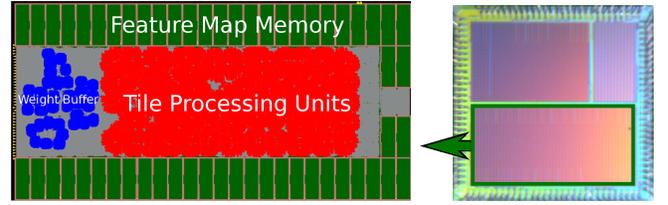

    \centering
    \includegraphics[width=.60\columnwidth]{\figures{floorplan3_labeled}} \includegraphics[width=.35\columnwidth]{\figures{chip_new}}
    \caption{Floorplan with Weight Buffer, Feature Map Memory and Tile Processing Units (left) and photograph of the taped-out multi-project chip Poseidon\textsuperscript{\ref{ref_poseideon}} with Hyperdrive on the bottom side.}
    \label{fig:floorplan}
\end{figure}

\footnotetext{Hyperdrive was taped-out alongside of two different projects (Kerbin and Quentin) on the same die to share costs, details can be found on \newline{\url{http://asic.ethz.ch/2018/Poseidon.html}\label{ref_poseideon}}}

Hyperdrive was designed in GF 22\,nm FDX technology using an 8 track low voltage threshold (LVT) standard cell library. This flavor of the technology allows to apply up to 1.8V of forward body biasing (FBB), increasing the operating frequency of the chip at the cost of higher leakage power. Synthesis was performed with Synopsys Design Compiler 2017.09, while place \& route was performed with Cadence Innovus 17.11.

The chip has an effective core area of 1.92\,mm\textsuperscript{2} (=9.6\,MGE)\footnote{One 2-input NAND gate equivalents (GE) is 0.199\,\textmu m\textsuperscript{2} in GF22.}, where 1.24\,mm\textsuperscript{2} are SRAM memories (6.4\,Mbit), 0.115\,mm\textsuperscript{2} are SCM memory (74\,kbit) and 0.32\,mm\textsuperscript{2} arithmetic units. \figref{fig:floorplan} shows on the right side a photograph of the actual chip and on the left side Hyperdrive's floorplan.

Testing and characterization (frequency, power) of silicon prototypes were performed on the industry-standard ASIC tester Advantest SoC V93000 and core power are based on the real chip measurements. The I/O energy was determined on the basis of an LPDDR3 PHY implemented in 28\,nm technology \cite{Cavigelli2016}, estimated as 21\,pJ/bit, as in context of our research no low-swing interface IP blocks were available. It should be noted that this has to be considered as quite optimistic bound for I/O energy in a low-cost chip (the LPDDR3 PHY is quite complex and expensive), hence pessimistic for the proposed architecture focusing on system-level energy efficiency and specifically I/O bandwidth reduction. If we use low-cost low-complexity full-swing I/O interfaces (used for the implementation of this prototype, and of the other state-of-the-art accelerator \cite{UNPU, Andri2016a, Wang2017, Han2016a}) would further magnify the system-level energy gain of Hyperdrive with respect to other architectures, but would probably give too much advantage to our solution with respect to industrialized, production-ready scenario where low-swing I/O interfaces would be used \cite{Jouppi2017}. 

\revONE{\figref{fig:energypie}} provides an overview of the Hyperdrive's blocks power consumption at the operating voltage of 0.5\,V and {58}\,MHz. The power consumption of memory arrays, memory periphery and logic were measured on the silicon prototype, available through the multi-power rails implementation strategy. On the other hand, the breakdown of the remaining standard cell logic power contributions is split into \tpus, Weight Buffer and Others and has been estimated with post-layout simulations. It is interesting to note that a considerable amount of the power is consumed into the arithmetic units, while only a small overhead comes from memory accesses and I/Os, due to the efficient exploitation of Feature Map stationary (i.e., temporal locality) of the Hyperdrive architecture, explaining its superior system-level energy efficiency with respect to the other BWN accelerators \revONE{in \tabref{tab:soa}}. The main features of the chip in other operating points is reported in \tabref{tab:metrics}

\begin{figure}
    \centering
    \includegraphics[width=1.0\columnwidth]{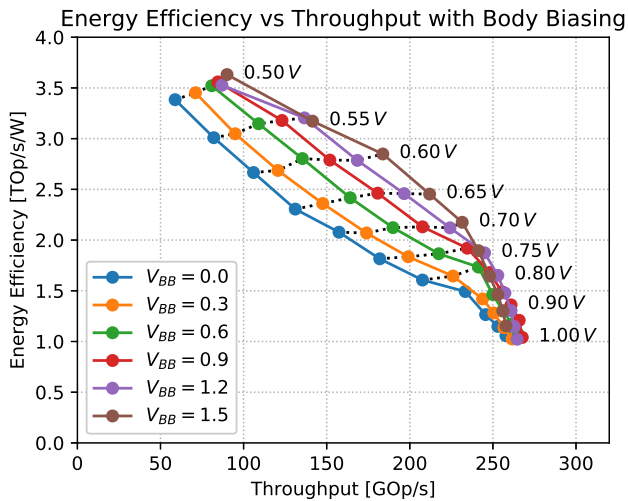}
    \caption{Energy Efficiency vs. Throughput for different Body Bias Voltages including I/O for ResNet-34.}
    \label{fig:theneffvbb}
\end{figure}

In order to characterize the best energy point of the chip we swept the body bias of the system along the available range (i.e., from 0\,V to 1.8\,V), as shown in \figref{fig:theneffvbb}. It is interesting to note that both performance and energy efficiency increase together with body biasing, due to the favorable ratio between leakage power and dynamic power (\markblue{4\%} at  0.5\,V with no body biasing) and that even if the memory arrays are not body biased (i.e., leakage does not increase) the operating frequency increases significantly. This makes the operating points at 1.5\,V FBB the most energy efficient ones for all performance targets. The best energy point occurs at 0.5\,V VDD and 1.5\,V FBB, featuring a throughput of 88\,TOp/s and an energy efficiency of \revONE{3.6}\,TOPS/W running ResNet-34.

\begin{figure}
    \centering
    \includegraphics[width=1\columnwidth]{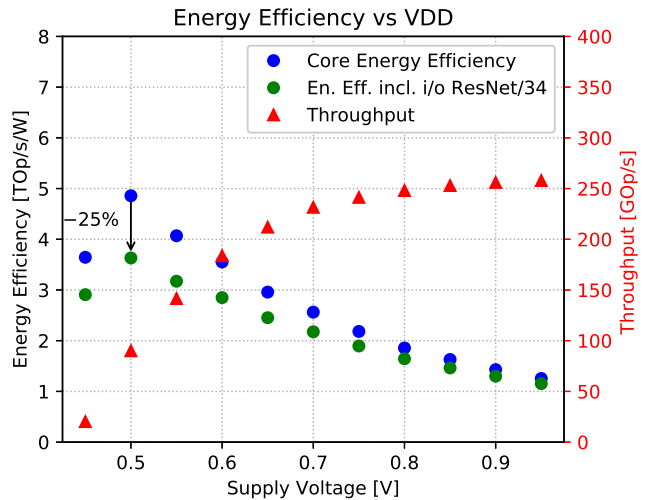}
    \caption{Energy Efficiency and Throughput vs. supply voltages}
    \label{fig:eneffthroughput}
\end{figure}

\figref{fig:eneffthroughput} shows the Energy Efficiency sweep vs. VDD. As mentioned before, the peak energy efficiency is achieved at 0.5V. Below this operating voltage, the relatively small operating frequency (i.e., 60\,MHz) makes the leakage dominate, hence efficiency drops. It is interesting to note that, as opposed to other architectures implemented in scaled technologies, where the IO energy is dominating \tabref{tab:soa}, in Hyperdrive the system level energy drops by only 25\% when introducing the I/O energy into the analysis.

\newcommand{\slice}[5]{
  \pgfmathparse{0.5*#1+0.5*#2}
  \let\midangle\pgfmathresult

  \draw[thick,fill=#5] (0,0) -- (#1:1) arc (#1:#2:1) -- cycle;

  \node[label=\midangle:#4] at (\midangle:1) {};

  \pgfmathparse{min((#2-#1-10)/110*(-0.3),0)}
  \let\temp\pgfmathresult
  \pgfmathparse{max(\temp,-0.5) + 0.8}
  \let\innerpos\pgfmathresult
  \node at (\midangle:\innerpos) {#3\%};
}
\begin{figure}
   
\begin{tikzpicture}[scale=2]

\newcounter{a}
\newcounter{b}
\foreach \p/\t/\c in { 1.0/WBuf (0.6\%)/blue, 17.5/FMM/OliveGreen!70, 
27.0/Tile PU (Arith.)/BrickRed!70,
22.4/\makecell{Tile PU \\(Acc. Registers)}/red!70,
27.9/{I/O}/yellow!70,
4.4/Others/black!30}
  {
    \setcounter{a}{\value{b}}
    \addtocounter{b}{\p}
    \slice{\thea/98*360}
          {\theb/98*360}
          {\p}{\t}{\c}
  }

\end{tikzpicture}
\caption{Ratio of energy consumption at 0.5\,V most energy-efficient corner.}\label{fig:energypie}
\end{figure}
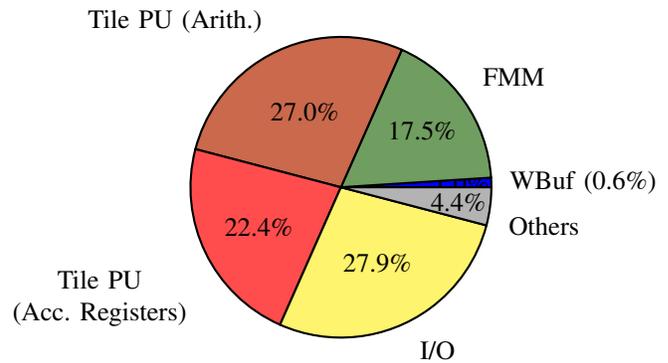
\begin{table}
\caption{Overview of Cycles, Throughput for ResNet-34}\label{tab:resneteval}
\centering
\begin{tabular}{lrrrr}
\toprule %
layer type & \#cycles & \#Op & \#Op/cycle & \#Op/s \\
\midrule
conv & 4.52\,M & 7.09\,G & 1568 & \\
bnorm & 59.90\,k & 2.94\,M & 49 & \\
bias & 59.90\,k & 2.94\,M & 49 & \\
bypass & 7.68\,k & 376.32\,k & 49 & \\
\midrule
total & 4.65\,M & 7.10\,G & 1.53\,k & 431\,G\\
\bottomrule
\end{tabular} 
\end{table} 

\newcolumntype{K}{>{\setbox0=\hbox\bgroup}c<{\egroup}@{}}
\begin{table}
\caption{{Overview of \textsc{Hyperdrive}} {(measured numbers)}}
\centering
\begin{tabular}{lrrrK}\toprule
Operating Point [V] & 0.5 & 0.65 & 0.8 & 0.9 \\ \midrule
Op. Frequency [MHz] & \calc{0}{57.42} & \calc{0}{135.15} & \calc{0}{158.2} & \calc{0}{163.28125} \\
Power [mW] & \calc{0}{21.57}   & \calc{0}{71.65484361} & \calc{0}{133.61318858} & \calc{0}{179.11946047} \\
Throughput [Op/cycle] & 1568 & 1568 & 1568 & 1568 \\
Throughput [GOp/s] & \calc{0}{1568/1000*57.42*0.975} & \calc{0}{1568/1000*135.15} & \calc{0}{1568/1000*158.2} & \calc{0}{1568/1000*163.28125} \\
Core Energy Eff. [TOp/s/W] & 4.9 & \calc{1}{1568/1000*135.15/71.65484361} & \calc{1}{1568/1000*158.2/133.61318858} & \calc{1}{1568/1000*163.28125/179.11946047} \\
Core Area [mm\textsuperscript{2}] & 1.92 & 1.92 & 1.92 & 1.92 \\
Memory [Mbit] & 6.4 & 6.4 & 6.4 & 6.4 \\\bottomrule
\end{tabular}   \label{tab:metrics}
\end{table}

\begin{table*}[t]
\centering
\caption{Comparison with State-of-the-Art BWN Accelerators (Top: Image Recognition, Bottom: Object Detection)}\label{tab:soa}
\begin{tabular}{l@{}llll@{\tableSoaSpacing}@{\tableSoaSpacing}r@{\tableSoaSpacing}r@{\tableSoaSpacing}r@{\tableSoaSpacing}r@{\tableSoaSpacing}r@{\tableSoaSpacing}r@{\tableSoaSpacing}r@{\tableSoaSpacing}r@{}}\toprule 
& Name & Techn. & DNN & \makecell{Input\\ Size} & \makecell{Precision\\ Wghts/Acts} & \makecell{Core\\\unit{V}} & \rmv{\makecell{Throughp.\\\unit{GOp/s}} &} \makecell{Eff. Th.\\\unit{GOp/s}} & \makecell{Core E\\\unit{mJ/im}} & \makecell{I/O E\\\unit{mJ/im}} & \makecell{Total E\\\unit{mJ/im}} & \makecell{En. Eff.\\\unit{TOp/s/W}} & \makecell{Area\\\unit{MGE}} \\\midrule 
 \multirow{10}{*}{\rotatebox{90}{\textit{Image Classification}}\hspace{2mm}} &   YodaNN (layout) \cite{Andri2016a} & umc65 & ResNet-34 & 224$^2$& Bin./Q12 & 1.20 & \rmv{1510 &} 490 & 0.9 & 3.6 & 4.5 & \calc{1}{7.22/4.5} & 1.3 \\ 
 &  YodaNN (layout) \cite{Andri2016a} & umc65 & ResNet-34 & 224$^2$& Bin./Q12 & 0.60 & \rmv{55 &} 18 & {0.1} & 3.6 & 3.7 &\calc{1}{7.22/3.7} & 1.3 \\ 

   &  \revONE{Wang w/ 25\,Mbit SRAM} & SMIC130 & ResNet-34 & 224$^2$& Bin./ENQ6 & 1.08 & \rmv{ &} 876 & 5.4 & 1.7 & 7.2 & 1.0 &  \\ 

   &  \revONE{UNPU (chip)} & 65\,nm & ResNet-34 & 224$^2$& Bin./Q16 & 0.77 & \rmv{?? &} 346 & \revONE{2.3} & \revONE{3.6} & \revONE{6.0} & \revONE{1.2} & 11.1 \\ 
  &  \jetcas{Hyperdrive (chip)} & GF22 & ResNet-34 & 224$^2$& Bin./FP16 & 0.50 & \rmv{88	&}88&	1.4&	0.5&	1.9&		3.6&	9.0 \\
  &  \jetcas{Hyperdrive (chip)} & GF22 & ResNet-34 & 224$^2$& Bin./FP16 & 1.00 & \rmv{ 268.3	&}	263	&6.5	&0.5	&7.0&	 \rmv{ 1.1	&}1.0 &	9.0 \\\cmidrule{2-13}
    
  &\revONE{Wang w/ 25\,Mbit SRAM} & SMIC130 &       ShuffleNet & 224$^2$& Bin./ENQ6&      1.08&   876 &   0.3 &   0.4 &   0.7      &   0.5      &9.9\\

&\revONE{UNPU (chip)} & 65\,nm & ShuffleNet & 224$^2$&              Bin./Q16&       0.77&                        \revONE{346}&    0.1 &   1.0 &   1.1      &      0.3     &       11.1\\

&\revONE{Hyperdrive (chip)} & GF22 & ShuffleNet & 224$^2$&           Bin./FP16&      0.50&   91&     0.1 &   0.1 &   0.2   &       2.1     &       9.0\\

\midrule

 \multirow{8}{*}{\rotatebox{90}{\textit{Object Detection}}\hspace{2mm}}

 &\revONE{Wang w/ 25\,Mbit SRAM} & SMIC130 &       YOLOv3(COCO) & 320$^2$& Bin./ENQ6&      1.08&   876 &   40.9 &  4.2 &   45.1     &   1.2      &9.9\\

&\revONE{UNPU (chip)} & 65\,nm & YOLOv3 & 320$^2$&           Bin./Q16&       0.77&                        \revONE{346}&    17.2 &  9.1 &   26.4     &      2.0     &       11.1\\

&\revONE{Hyperdrive (chip)} & GF22 & YOLOv3 & 320$^2$&           Bin./FP16&      0.50&   75&     13.1 &  1.4 &   14.5  &       3.7     &       9.0\\
\cmidrule{2-13}

&\jetcas{Wang w/ 25\,Mbit SRAM}&	SMIC130&	ResNet-34&	2k\x 1k& Bin/ENQ6& 	{}		&			{}	&243.4	&40.5&	283.9	& \calc{1}{293.2/283.9} &		{} \\

&\jetcas{UNPU (chip) \cite{UNPU}} & 65\,nm & ResNet-34 & 2k\x 1k & Bin./Q16 & 0.77 & \rmv{?? &} \revONE{346} & 97.7 & \revONE{105.6} & \revONE{\calc{1}{97.7+105.6}} &\revONE{\calc{1}{293.2/(97.7+105.6)}} & 11.1 \\

  &  \jetcas{Hyperdrive (10\x 5)} & GF22 & ResNet-34 & 2k\x 1k & Bin./FP16 & 0.50 &  \rmv{1960.0 &} 4547 &	61.9	&7.6&	69.5&	 \rmv{6.3&}	4.3& 50\x 9.0 \\\cmidrule{2-13}
    
  &  \jetcas{Hyperdrive (20\x 10)} & GF22 & ResNet-152 & 2k\x 1k & Bin./FP16 & 0.50 &   \rmv{ 7840&}	18189 &	185.2&	21.6&	206.8&	\rmv{6.3&}	4.4&	200\x 9.0 \\

\midrule
\ifdefined\wangold
\multicolumn{8}{r}{ \textcolor{red}{Improvement over state-of-the-art for object detection:}} \rmv{&  \textcolor{red}{0.}33\x} &  \textcolor{red}{61\x} &  \textcolor{red}{2.1\x} &  \textcolor{red}{2.1\x} & \\ 
\fi

\multicolumn{9}{r}{\rv{Improvement over state-of-the-art for image classification \revONE{(ResNet-34)}:}} \rmv{& \calc{1}{0.1/1.1}\x} & \calc{1}{1.7/0.49}\x & \calc{1}{3.6/2}\x & \calc{1}{3.6/2}\x & \\ 
\multicolumn{9}{r}{\rv{Improvement over state-of-the-art for object detection: \revONE{(ResNet-34)}:}}  \rmv{& \calc{1}{0.1/1.1}\x} & \calc{1}{40.5/7.6}\x & \calc{1}{4.3/1.4}\x & \calc{1}{4.3/1.4}\x & \\ 

\cmidrule{9-12}
\end{tabular}

\end{table*}

\subsection{Benchmarking}\label{sec:benchmarking}

\TBC{The main evaluation of Hyperdrive has been performed on ResNet-34, whose network structure have been used in plenty of applications.} This network features a good trade-off between depth and accuracy, i.e., ResNet-50 outperforms ResNet-34 by just 0.5\% (Top-1) in terms of classification accuracy on the ImageNet dataset, but is roughly 50\% more compute-intensive and the memory footprint is even \calc{1}{20.8/6.4}$\times$ higher (see \secref{sec:systolicsystemscaling}). 

The first and the last layer need to stay in full-precision to keep a satisfactory accuracy and are not implemented on Hyperdrive, but they contribute just 3\% of the computation (226\,MOp of 7.3\,GOp) and can therefore also be evaluated on low-power compute platforms \cite{Gautschi2017}. 

\tabref{tab:resneteval} provides an overview of the number of operations, number of cycles and throughput while Hyperdrive is evaluating ResNet-34. In case of batch normalization, the throughput is reduced since just 49 multipliers are available and the normalization does take more cycles. In the layers where the bypass has to be added, Hyperdrive can also just calculate one output FM at a time, because the memory bandwidth is limited to 49 half-precision words. Fortunately, the non-convolution operations are comparably rare and a real throughput of 1.53\,kOp/cycle or 221.9\,GOp/s @\,0.65\,V is achieved leading to a very high utilization ratio of \calc{1}{100/1.57*1.53}\% of the peak throughput. \revONE{\tabref{tab:utilization} provides an overview of the utilization (i.e., actual throughput normalized to theoretical peak throughput) for several networks. It can be seen that both ResNet-34 and ShuffleNet have very high utilization since the feature maps tile equally onto the \tpus{}. In the other case, where the intermediate feature maps are not sized by a multiple of M\x N (i.e., 7\x7), the feature maps are padded with zeros and the last row and column of \tpus{} is idle during calculation of these zero pixels. Nevertheless, also in these cases, utilization is well above 80\% (e.g., YOLOv3 \cite{redmon2018yolov3} on a 320\x 320 with 82.8\%), which confirms the high flexibility of the proposed architecture with respect to different flavors of network topologies.}

\begin{table}
\revONE{
\caption{Utilization of Hyperdrive}\label{tab:utilization}
\centering
\begin{tabular}{lrrrr}
\toprule %
Network (Resolution) & \#Op & \#cycles & \#Op/cycle & Utilization \\
\midrule
Baseline (Peak Perf.) &   &   & 1.57\,k & 100.0\%\\
ResNet-34 (224$^2$) & 7.10\,G & 4.65\,M & 1.53\,k & 97.5\%\\
ShuffleNet (224$^2$) & 140\,M & 90.3\,k & 1.55\,k & 98.8\% \\
YOLOv3 (320$^2$) & 53.1\,G & 33.9\,M & 1.30\,k & 82.8\% \\
\bottomrule
\end{tabular} 
}
\end{table} 

\subsection{I/O in Multi-Chip Setup}\label{sec:ioconsider}
\begin{figure}
    \centering
    \includegraphics[width=.9\columnwidth]{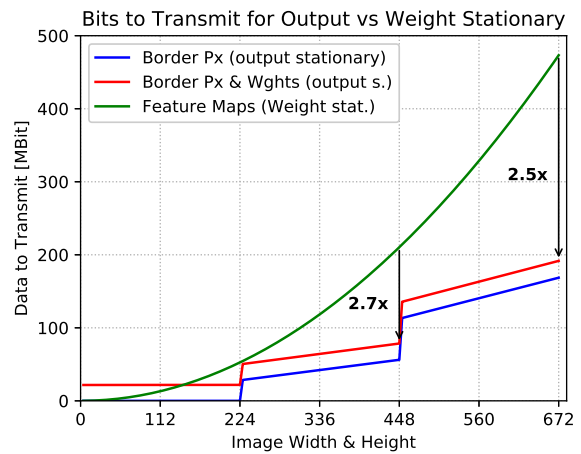}
    \caption{\revONE{Number of bits to be transmitted with the weight stationary approach compared to the output stationary approach adopted in the Hyperdrive architecture (including border exchange).}}
    \label{fig:bits2transmit}
\end{figure}
Having multiple-chips introduces implicitly more I/O as the border pixels have to be sent to the neighboring chips. To illustrate the relation between the feature map size to the amount of I/O, \figref{fig:bits2transmit} compares the common weight stationary approach (green) to the feature map stationary approach of Hyperdrive (red). The evaluation is done with ResNet-34 with the taped-out accelerator dimensioned to fit the WCL for 3\x 224\x 224 sized images. By scaling the spatial dimensions evenly, the amount of I/O stays constant for the weights of 21.6\,Mbit until the maximum dimension of 224\x224 is reached. After that the FM is tiled onto several chips, starting with 2\x 2. This introduces the need exchange two entire rows and columns per output channel and layer to transmit and increases linearly with the FM size until the FM is not fitting anymore onto the 2\x2 chips, and tiling is done on 3\x3, etc.
In case of a systolic array of 2\x2 chips, the I/O can be reduced by up to 2.7\x{} and 2.5\x{} for a 3\x3 array while accounting for the border exchanges.

\subsection{Comparison with State-of-the-Art}\label{sec:compsoa}
\revONE{\tabref{tab:soa} compares Hyperdrive with state-of-the-art binary weight CNN accelerators. The upper part of the table compares the SoA accelerators running image recognition applications (i.e., ResNet-34, VGG-16 and ShuffleNet on 224\x 224 sized images), while the lower part compares key metrics coming from object detection applications with images available in autonomous driving data sets \cite{Cordts2016, Wu2016a} (i.e., ResNet-34 on 2048\x 1024, and YOLOv3 on 320\x{}320 images). \rev{At 0.65\,V, Hyperdrive achieves a frame rate of \markblue{46.7} for ResNet-34}, and, most important, the performance is independent of the image resolution thanks to the systolically scalable capabilities of the architecture.}

\revONE{While the totality of previous works is dominated by I/O energy, especially for spatially large feature maps, in Hyperdrive the I/O energy is only a small factor of the total energy (7\% to 30\%, depending on the application). Thanks to this feature, Hyperdrive outperforms other architectures by up to 1.8\x{} on image classification applications and up to 3.1\x{} in object detection applications, in terms of energy efficiency. More precisely, if we compare with the architecture presented in \cite{Wang2017}, Hyperdrive is 3.1\x{} more energy efficient, despite the number of bits used for the FMs in ENQ6 is only 6 \cite{Wang2017}, hence higher energy efficiency is achieved with much less aggressive reduction of HW precision.}
\revONE{It should also be mentioned here, that previous work has estimated that for equi-precision results, highly discretized networks need to be just slightly larger (e.g., a ternary-weight (Q2) ResNet-18 is about 12\% larger than a full-precision GoogLeNet while both achieve the same accuracy when trained with precision-aware algorithms \cite{AojunZhou2016}), whereas the core energy efficiency would improve significantly from stronger quantization and therefore Hyperdrive is expected to outperform the state-of-the-art even more than the 3.1\x{} factor reported here when using fixed-point representation and stronger quantization.}

\revONE{Furthermore, we compare our work with UNPU \cite{UNPU}, which is the only silicon implementation adopting fixed-point arithmetic with adaptable precision (16, 8, 4, 2, 1) for the feature maps. We compare with the 16-bit mode, as this is the most similar with respect to accuracy. Our approach uses up to \markblue{}{5.3}\x{} less energy for I/O and increases overall energy efficiency by up to \markblue{3}\x{} since just the first input FM and the weights need to be streamed to the chip, but not the intermediate FMs. ShuffleNet is a challenging task for all the three accelerators analyzed, as the feature maps are very deep, but spatially small. This implies a low compute intensity relative to the number of weights, which is an adverse pattern for Hyperdrive, and for most accelerators. On the other hand, grouping implies that for every group of output channels, just the subset of assigned input channels is filtered, which reduces the compute complexity while keeping the same feature map volume and is therefore an aspect in Hyperdrive's favor. Thus Hyperdrive still outperforms the state-of-the-art by 4.2\x.}

\revONE{The previous state-of-the-art accelerators are designed in less advanced technologies than Hyperdrive (GF\,22nm compared to 65\,nm and 130\,nm), thus their core energy efficiency would be improved by using an advanced technology. Nevertheless, Hyperdrive's core energy efficiency is \calc{1}{60/4.9}\x{} worse than YodaNN's and just 1.6 or 3.7\x{} better than UNPU and Wang \emph{et al.}} One of the reasons is that we use FP16 operators which are more robust than Q12 or ENQ6 in \cite{Andri2016a,Wang2017} and were shown to work with the most challenging deep networks. Using floating-point feature maps directly impacts the energy for the accumulation operations as well as memory and register read/write operations. \rv{ENQ on the other side has been shown to introduce an accuracy drop of 1.6\% already on CIFAR-100 \cite{Wang2017}, which is more than the difference between running ResNet-34 instead of ResNet-110 on CIFAR-10. It thus implies that a deeper network has to be computed to achieve a comparable accuracy.}
\revONE{Furthermore, optimizations such as approximate adders and strong quantization have not been implemented, but can be combined with Hyperdrive's concepts, coupling core efficiency gains with the removal of the non-scalable I/O bottleneck. For instance, moving from FP16 to Q12 would lead to an energy efficiency boost that can be estimated  to be around 3\x{} for the core, which would translate to a system efficiency boost of \markblue{6.8\x{}} for high accuracy object detection with ResNet-34 features. }

\section{Conclusion}
\label{sec:conclusion}
We have presented Hyperdrive: a systolically scalable hardware architecture for binary-weight neural networks, which dramatically minimizes the I/O energy consumption to achieve outstanding system-level energy efficiency. Hyperdrive achieves an energy efficiency of \revONE{4.3}\,TOp/s/W on object detection task which is more than \revTWO{3.1}\x{} better than prior state-of-the-art architectures, by exploiting a binary-weight streaming mechanism while keeping the entire FMs on-chip. Furthermore, while previous architectures were limited to some specific network sizes, Hyperdrive allows running networks not fitting on a single die, by arranging multiple chips in an on-board 2D systolic array, scaling-up the resolution of neural networks, hence enabling a new class of applications such as object detection on the edge of the IoT.

\section*{Acknowledgements}
This project was supported in part by the Swiss National Science Foundation under grant no. 162524 (MicroLearn: Micropower Deep Learning), armasuisse Science \& Technology and the EU's H2020 program under grant no. 732631 (OPRECOMP).

{


\bibliographystyle{refsBst}
}
\vspace{-1.5cm}
\begin{IEEEbiography}[{\includegraphics[width=1in,height=1.25in,clip,keepaspectratio]{\figures{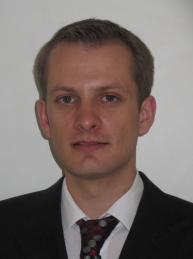}}}]{Renzo Andri}  received the M.Sc. degree in electrical engineering and information technology from ETH Zurich, Zurich, Switzerland, in 2015. He is currently pursuing a Ph.D. degree at the Integrated System Laboratory, ETH Zurich. His main research interests involve the design of low-power hardware accelerators for machine learning applications, and studying new algorithmic methods to further increase the energy-efficiency and therefore the usability of ML on energy-restricted devices.
\vspace{-1.5cm} 
\end{IEEEbiography}
\begin{IEEEbiography}[{\includegraphics[width=1in,height=1.25in,clip,keepaspectratio]{\figures{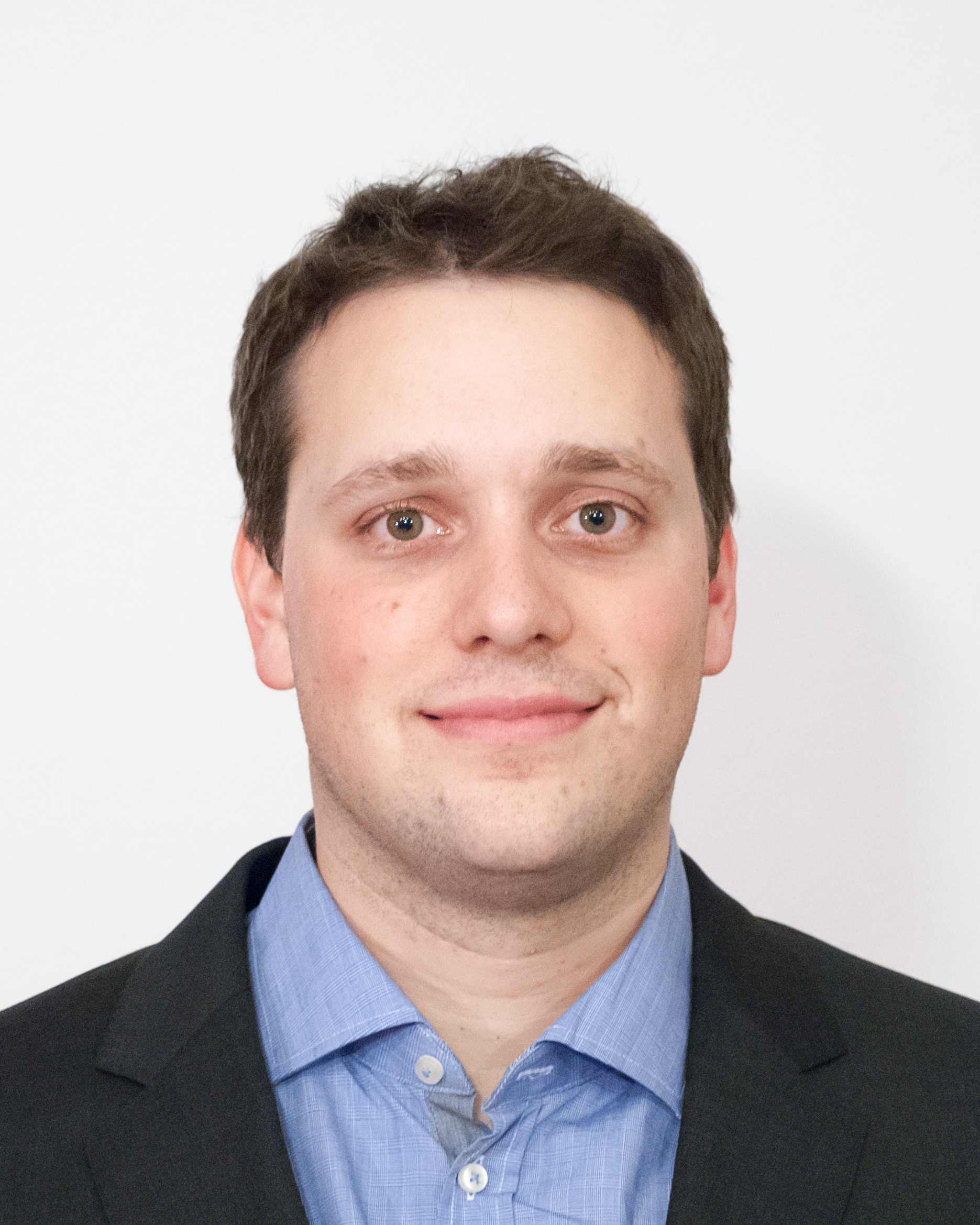}}}]{Lukas Cavigelli} received the B.Sc., M.Sc., and Ph.D. degree in electrical engineering and information technology from ETH Z\"urich, Z\"urich, Switzerland in 2012, 2014 and 2019, respectively. He has since been a postdoctoral researcher with ETH Z\"urich. 
His research interests include deep learning, computer vision, embedded systems, and low-power integrated circuit design. He has received the best paper award at the VLSI-SoC and the ICDSC conferences in 2013 and 2017, and the best student paper award at the Security+Defense conference in 2016. 
\vspace{-1.5cm}
\end{IEEEbiography}
\begin{IEEEbiography}[{\includegraphics[width=1in,height=1.25in,clip,keepaspectratio]{\figures{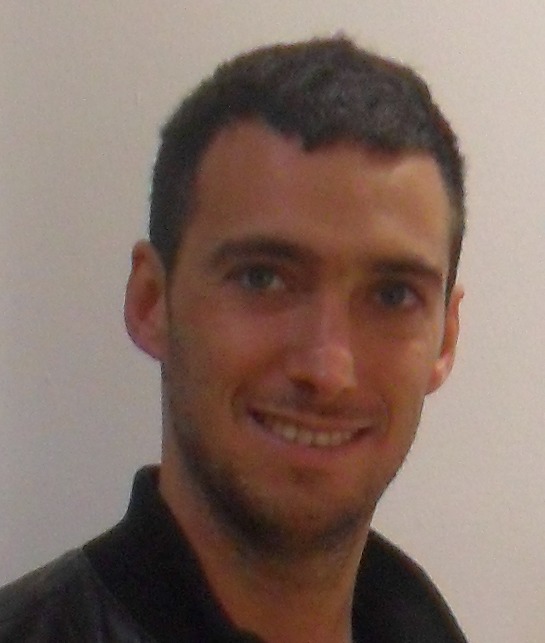}}}]{Davide Rossi} received the Ph.D. from the University of Bologna, Italy, in 2012. He has been a post doc researcher in the Department of Electrical, Electronic and Information Engineering “Guglielmo Marconi” at the University of Bologna since 2015, where he currently holds an assistant professor position. His research interests focus on energy-efficient digital architectures in the domain of heterogeneous and reconfigurable multi- and many-core systems on a chip. This includes architectures, design implementation strategies, and run-time support to address performance, energy efficiency, and reliability issues of both high end embedded platforms and ultra-low-power computing platforms targeting the IoT domain. In these fields, he has published more than 80 papers in international peer-reviewed conferences and journals.
\vspace{-1.5cm}
\end{IEEEbiography}
\begin{IEEEbiography}[{\includegraphics[width=1in,height=1.2in,clip,keepaspectratio]{\figures{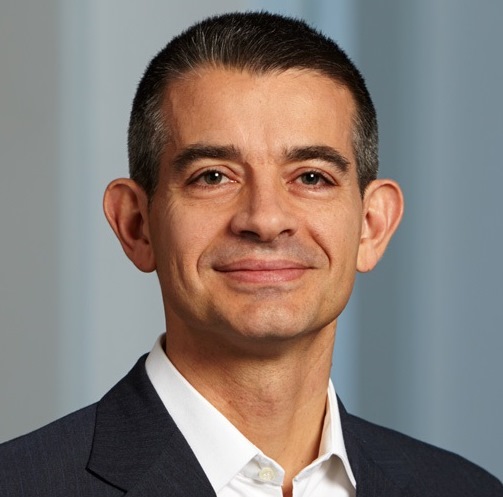}}}]{Luca Benini}  is the Chair of Digital Circuits and Systems at ETH Zurich and a Full Professor at the University of Bologna. He has served as Chief Architect for the Platform2012 in STMicroelectronics, Grenoble. Dr. Benini's research interests are in energy-efficient system and multi-core SoC design.  He is also active in the area of energy-efficient smart sensors and sensor networks. 
He has published more than 1'000 papers in peer-reviewed international journals and conferences, four books and several book chapters. He is a Fellow of the ACM and of the IEEE and a member of the Academia Europaea. 
\vspace{-1.5cm}
\end{IEEEbiography}

\end{document}